# A Study of the Energy Dependence of the Underlying Event in Proton-Antiproton Collisions

T. Aaltonen,[21] M. Albrow,[15] S. Amerio[ll,39] D. Amidei,[31] A. Anastassov[w,15] A. Annovi,[17] J. Antos,[12] G. Apollinari,[15] J.A. Appel,[15] T. Arisawa,[52] A. Artikov,[13] J. Asaadi,[47] W. Ashmanskas,[15] B. Auerbach,[2] A. Aurisano,[47] F. Azfar,[38] W. Badgett,[15] T. Bae,[25] A. Barbaro-Galtieri,[26] V.E. Barnes,[43] B.A. Barnett,[23] P. Barria[nn,41] P. Bartos,[12] M. Bauce[ll,39] F. Bedeschi,[41] S. Behari,[15] G. Bellettini[mm,41] J. Bellinger,[54] D. Benjamin,[14] A. Beretvas,[15] A. Bhatti,[45] K.R. Bland,[5] B. Blumenfeld,[23] A. Bocci,[14] A. Bodek,[44] D. Bortoletto,[43] J. Boudreau,[42] A. Boveia,[11] L. Brigliadori[kk,6] C. Bromberg,[32] E. Brucken,[21] J. Budagov,[13] H.S. Budd,[44] K. Burkett,[15] G. Busetto[ll,39] P. Bussey,[19] P. Butti[mm,41] A. Buzatu,[19] A. Calamba,[10] S. Camarda,[4] M. Campanelli,[28] F. Canelli[ee,11] B. Carls,[22] D. Carlsmith,[54] R. Carosi,[41] S. Carrillo[l,16] B. Casal[j,9] M. Casarsa,[48] A. Castro[kk,6] P. Catastini,[20] D. Cauz[sstt,48] V. Cavaliere,[22] A. Cerri[e,26] L. Cerrito[r,28] Y.C. Chen,[1] M. Chertok,[7] G. Chiarelli,[41] G. Chlachidze,[15] K. Cho,[25] D. Chokheli,[13] A. Clark,[18] C. Clarke,[53] M.E. Convery,[15] J. Conway,[7] M. Corbo[z,15] M. Cordelli,[17] C.A. Cox,[7] D.J. Cox,[7] M. Cremonesi,[41] D. Cruz,[47] J. Cuevas[y,9] R. Culbertson,[15] N. d'Ascenzo[v,15] M. Datta[hh,15] P. de Barbaro,[44] L. Demortier,[45] M. Deninno,[6] M. D'Errico[ll,39] F. Devoto,[21] A. Di Canto[mm,41] B. Di Ruzza[p,15] J.R. Dittmann,[5] S. Donati[mm,41] M. D'Onofrio,[27] M. Dorigo[uu,48] A. Driutti[sstt,48] K. Ebina,[52] R. Edgar,[31] A. Elagin,[47] R. Erbacher,[7] S. Errede,[22] B. Esham,[22] S. Farrington,[38] J.P. Fernández Ramos,[29] R. Field,[16] G. Flanagan[t,15] R. Forrest,[7] M. Franklin,[20] J.C. Freeman,[15] H. Frisch,[11] Y. Funakoshi,[52] C. Galloni[mm,41] A.F. Garfinkel,[43] P. Garosi[nn,41] H. Gerberich,[22] E. Gerchtein,[15] S. Giagu,[46] V. Giakoumopoulou,[3] K. Gibson,[42] C.M. Ginsburg,[15] N. Giokaris,[3] P. Giromini,[17] V. Glagolev,[13] D. Glenzinski,[15] M. Gold,[34] D. Goldin,[47] A. Golossanov,[15] G. Gomez,[9] G. Gomez-Ceballos,[30] M. Goncharov,[30] O. González López,[29] I. Gorelov,[34] A.T. Goshaw,[14] K. Goulianos,[45] E. Gramellini,[6] C. Grosso-Pilcher,[11] R.C. Group,[51,15] J. Guimaraes da Costa,[20] S.R. Hahn,[15] J.Y. Han,[44] F. Happacher,[17] K. Hara,[49] M. Hare,[50] R.F. Harr,[53] T. Harrington-Taber[m,15] K. Hatakeyama,[5] C. Hays,[38] J. Heinrich,[40] M. Herndon,[54] A. Hocker,[15] Z. Hong,[47] W. Hopkins[f,15] S. Hou,[1] R.E. Hughes,[35] U. Husemann,[55] M. Hussein[cc,32] J. Huston,[32] G. Introzzi[ppqq,41] M. Iori[rr,46] A. Ivanov[o,7] E. James,[15] D. Jang,[10] B. Jayatilaka,[15] E.J. Jeon,[25] S. Jindariani,[15] M. Jones,[43] K.K. Joo,[25] S.Y. Jun,[10] T.R. Junk,[15] M. Kambeitz,[24] T. Kamon,[25,47] P.E. Karchin,[53] A. Kasmi,[5] Y. Kato[n,37] W. Ketchum[ii,11] J. Keung,[40] B. Kilminster[ee,15] D.H. Kim,[25] H.S. Kim[bb,15] J.E. Kim,[25] M.J. Kim,[17] S.H. Kim,[49] S.B. Kim,[25] Y.J. Kim,[25] Y.K. Kim,[11] N. Kimura,[52] M. Kirby,[15] K. Knoepfel,[15] K. Kondo,[52,*] D.J. Kong,[25] J. Konigsberg,[16] A.V. Kotwal,[14] M. Kreps,[24] J. Kroll,[40] M. Kruse,[14] T. Kuhr,[24] M.

Kurata,[49] A.T. Laasanen,[43] S. Lammel,[15] M. Lancaster,[28] K. Lannon[x],[35] G. Latino[nn],[41] H.S. Lee,[25] J.S. Lee,[25] S. Leo,[22] S. Leone,[41] J.D. Lewis,[15] A. Limosani[s],[14] E. Lipeles,[40] A. Lister[a],[18] H. Liu,[51] Q. Liu,[43] T. Liu,[15] S. Lockwitz,[55] A. Loginov,[55] D. Lucchesi[ll],[39] A. Lucá,[17] J. Lueck,[24] P. Lujan,[26] P. Lukens,[15] G. Lungu,[45] J. Lys,[26] R. Lysak[d],[12] R. Madrak,[15] P. Maestro[nn],[41] S. Malik,[45] G. Manca[b],[27] A. Manousakis-Katsikakis,[3] L. Marchese[jj],[6] F. Margaroli,[46] P. Marino[oo],[41] K. Matera,[22] M.E. Mattson,[53] A. Mazzacane,[15] P. Mazzanti,[6] R. McNulty[i],[27] A. Mehta,[27] P. Mehtala,[21] C. Mesropian,[45] T. Miao,[15] D. Mietlicki,[31] A. Mitra,[1] H. Miyake,[49] S. Moed,[15] N. Moggi,[6] C.S. Moon[z],[15] R. Moore[ff gg],[15] M.J. Morello[oo],[41] A. Mukherjee,[15] Th. Muller,[24] P. Murat,[15] M. Mussini[kk],[6] J. Nachtman[m],[15] Y. Nagai,[49] J. Naganoma,[52] I. Nakano,[36] A. Napier,[50] J. Nett,[47] C. Neu,[51] T. Nigmanov,[42] L. Nodulman,[2] S.Y. Noh,[25] O. Norniella,[22] L. Oakes,[38] S.H. Oh,[14] Y.D. Oh,[25] I. Oksuzian,[51] T. Okusawa,[37] R. Orava,[21] L. Ortolan,[4] C. Pagliarone,[48] E. Palencia[e],[9] P. Palni,[34] V. Papadimitriou,[15] W. Parker,[54] G. Pauletta[sstt],[48] M. Paulini,[10] C. Paus,[30] T.J. Phillips,[14] G. Piacentino[q],[15] E. Pianori,[40] J. Pilot,[7] K. Pitts,[22] C. Plager,[8] L. Pondrom,[54] S. Poprocki[f],[15] K. Potamianos,[26] A. Pranko,[26] F. Prokoshin[aa],[13] F. Ptohos[g],[17] G. Punzi[mm],[41] I. Redondo Fernández,[29] P. Renton,[38] M. Rescigno,[46] F. Rimondi,[6,*] L. Ristori,[41,15] A. Robson,[19] T. Rodriguez,[40] S. Rolli[h],[50] M. Ronzani[mm],[41] R. Roser,[15] J.L. Rosner,[11] F. Ruffini[nn],[41] A. Ruiz,[9] J. Russ,[10] V. Rusu,[15] W.K. Sakumoto,[44] Y. Sakurai,[52] L. Santi[sstt],[48] K. Sato,[49] V. Saveliev[v],[15] A. Savoy-Navarro[z],[15] P. Schlabach,[15] E.E. Schmidt,[15] T. Schwarz,[31] L. Scodellaro,[9] F. Scuri,[41] S. Seidel,[34] Y. Seiya,[37] A. Semenov,[13] F. Sforza[mm],[41] S.Z. Shalhout,[7] T. Shears,[27] P.F. Shepard,[42] M. Shimojima[u],[49] M. Shochet,[11] I. Shreyber-Tecker,[33] A. Simonenko,[13] K. Sliwa,[50] J.R. Smith,[7] F.D. Snider,[15] H. Song,[42] V. Sorin,[4] R. St. Denis,[19,*] M. Stancari,[15] D. Stentz[w],[15] J. Strologas,[34] Y. Sudo,[49] A. Sukhanov,[15] I. Suslov,[13] K. Takemasa,[49] Y. Takeuchi,[49] J. Tang,[11] M. Tecchio,[31] P.K. Teng,[1] J. Thom[f],[15] E. Thomson,[40] V. Thukral,[47] D. Toback,[47] S. Tokar,[12] K. Tollefson,[32] T. Tomura,[49] D. Tonelli[e],[15] S. Torre,[17] D. Torretta,[15] P. Totaro,[39] M. Trovato[oo],[41] F. Ukegawa,[49] S. Uozumi,[25] F. Vázquez[l],[16] G. Velev,[15] C. Vellidis,[15] C. Vernieri[oo],[41] M. Vidal,[43] R. Vilar,[9] J. Vizán[dd],[9] M. Vogel,[34] G. Volpi,[17] P. Wagner,[40] R. Wallny[j],[15] S.M. Wang,[1] D. Waters,[28] W.C. Wester III,[15] D. Whiteson[c],[40] A.B. Wicklund,[2] S. Wilbur,[7] H.H. Williams,[40] J.S. Wilson,[31] P. Wilson,[15] B.L. Winer,[35] P. Wittich[f],[15] S. Wolbers,[15] H. Wolfe,[35] T. Wright,[31] X. Wu,[18] Z. Wu,[5] K. Yamamoto,[37] D. Yamato,[37] T. Yang,[15] U.K. Yang,[25] Y.C. Yang,[25] W.-M. Yao,[26] G.P. Yeh,[15] K. Yi[m],[15] J. Yoh,[15] K. Yorita,[52] T. Yoshida[k],[37] G.B. Yu,[14] I. Yu,[25] A.M. Zanetti,[48] Y. Zeng,[14] C. Zhou,[14] and S. Zucchelli[kk6]

(CDF Collaboration), †

[1]Institute of Physics, Academia Sinica, Taipei, Taiwan 11529, Republic of China
[2]Argonne National Laboratory, Argonne, Illinois 60439, USA
[3]University of Athens, 157 71 Athens, Greece

[4]Institut de Fisica d'Altes Energies, ICREA,
Universitat Autonoma de Barcelona, E-08193, Bellaterra (Barcelona), Spain
[5]Baylor University, Waco, Texas 76798, USA
[6]Istituto Nazionale di Fisica Nucleare Bologna,
[kk]University of Bologna, I-40127 Bologna, Italy
[7]University of California, Davis, Davis, California 95616, USA
[8]University of California, Los Angeles, Los Angeles, California 90024, USA
[9]Instituto de Fisica de Cantabria, CSIC-University of Cantabria, 39005 Santander, Spain
[10]Carnegie Mellon University, Pittsburgh, Pennsylvania 15213, USA
[11]Enrico Fermi Institute, University of Chicago, Chicago, Illinois 60637, USA
[12]Comenius University, 842 48 Bratislava,
Slovakia; Institute of Experimental Physics, 040 01 Kosice, Slovakia
[13]Joint Institute for Nuclear Research, RU-141980 Dubna, Russia
[14]Duke University, Durham, North Carolina 27708, USA
[15]Fermi National Accelerator Laboratory, Batavia, Illinois 60510, USA
[16]University of Florida, Gainesville, Florida 32611, USA
[17]Laboratori Nazionali di Frascati, Istituto Nazionale di Fisica Nucleare, I-00044 Frascati, Italy
[18]University of Geneva, CH-1211 Geneva 4, Switzerland
[19]Glasgow University, Glasgow G12 8QQ, United Kingdom
[20]Harvard University, Cambridge, Massachusetts 02138, USA
[21]Division of High Energy Physics, Department of Physics,
University of Helsinki, FIN-00014, Helsinki,
Finland; Helsinki Institute of Physics, FIN-00014, Helsinki, Finland
[22]University of Illinois, Urbana, Illinois 61801, USA
[23]The Johns Hopkins University, Baltimore, Maryland 21218, USA
[24]Institut für Experimentelle Kernphysik, Karlsruhe Institute of Technology, D-76131 Karlsruhe, Germany
[25]Center for High Energy Physics: Kyungpook National University,
Daegu 702-701, Korea; Seoul National University, Seoul 151-742,
Korea; Sungkyunkwan University, Suwon 440-746,
Korea; Korea Institute of Science and Technology Information,
Daejeon 305-806, Korea; Chonnam National University,
Gwangju 500-757, Korea; Chonbuk National University, Jeonju 561-756,
Korea; Ewha Womans University, Seoul, 120-750, Korea
[26]Ernest Orlando Lawrence Berkeley National Laboratory, Berkeley, California 94720, USA
[27]University of Liverpool, Liverpool L69 7ZE, United Kingdom
[28]University College London, London WC1E 6BT, United Kingdom
[29]Centro de Investigaciones Energeticas Medioambientales y Tecnologicas, E-28040 Madrid, Spain
[30]Massachusetts Institute of Technology, Cambridge, Massachusetts 02139, USA
[31]University of Michigan, Ann Arbor, Michigan 48109, USA
[32]Michigan State University, East Lansing, Michigan 48824, USA
[33]Institution for Theoretical and Experimental Physics, ITEP, Moscow 117259, Russia
[34]University of New Mexico, Albuquerque, New Mexico 87131, USA
[35]The Ohio State University, Columbus, Ohio 43210, USA
[36]Okayama University, Okayama 700-8530, Japan
[37]Osaka City University, Osaka 558-8585, Japan
[38]University of Oxford, Oxford OX1 3RH, United Kingdom
[39]Istituto Nazionale di Fisica Nucleare, Sezione di Padova,
[ll]University of Padova, I-35131 Padova, Italy
[40]University of Pennsylvania, Philadelphia, Pennsylvania 19104, USA
[41]Istituto Nazionale di Fisica Nucleare Pisa,
[mm]University of Pisa, [nn]University of Siena,
[oo]Scuola Normale Superiore, I-56127 Pisa,
Italy, [pp]INFN Pavia, I-27100 Pavia, Italy,
[qq]University of Pavia, I-27100 Pavia, Italy
[42]University of Pittsburgh, Pittsburgh, Pennsylvania 15260, USA
[43]Purdue University, West Lafayette, Indiana 47907, USA
[44]University of Rochester, Rochester, New York 14627, USA
[45]The Rockefeller University, New York, New York 10065, USA
[46]Istituto Nazionale di Fisica Nucleare, Sezione di Roma 1,
[rr]Sapienza Università di Roma, I-00185 Roma, Italy
[47]Mitchell Institute for Fundamental Physics and Astronomy,
Texas A&M University, College Station, Texas 77843, USA

[48]Istituto Nazionale di Fisica Nucleare Trieste,
[ss]Gruppo Collegato di Udine, [tt]University of Udine, I-33100 Udine,
Italy, [uu]University of Trieste, I-34127 Trieste, Italy
[49]University of Tsukuba, Tsukuba, Ibaraki 305, Japan
[50]Tufts University, Medford, Massachusetts 02155, USA
[51]University of Virginia, Charlottesville, Virginia 22906, USA
[52]Waseda University, Tokyo 169, Japan
[53]Wayne State University, Detroit, Michigan 48201, USA
[54]University of Wisconsin, Madison, Wisconsin 53706, USA
[55]Yale University, New Haven, Connecticut 06520, USA

[†]With visitors from [a]University of British Columbia, Vancouver, BC V6T 1Z1, Canada, [b]Istituto Nazionale di Van-Fisica Nucleare, Sezione di Cagliari, 09042 Monserrato (Cagliari), Italy, [c]University of California Irvine, Irvine, CA 92697, USA, [d]Institute of Physics, Academy of Sciences of the Czech Republic, 182 21, Czech Republic, [e]CERN, CH-1211 Geneva, Switzerland, [f]Cornell University, Ithaca, NY 14853, USA, [g]University of Cyprus, Nicosia CY-1678, Cyprus, [h]Office of Science, U.S. Department of Energy, Washington, DC 20585, USA, [i]University College Dublin, Dublin 4, Ireland, [j]ETH, 8092 Zürich, Switzerland, [k]University of Fukui, Fukui City, Fukui Prefecture, Japan 910-0017, [l]Universidad Iberoamericana, Lomas de Santa Fe, México, C.P. 01219, Distrito Federal, [m]University of Iowa, Iowa City, IA 52242, USA, [n]Kinki University, Higashi-Osaka City, Japan 577-8502, [o]Kansas State University, Manhattan, KS 66506, USA, [p]Brookhaven National Laboratory, Upton, NY 11973, USA, [q]Istituto Nazionale di Fisica Nucleare, Sezione di Lecce, Via Arnesano, I-73100 Lecce, Italy, [r]Queen Mary, University of London, London, E1 4NS, United Kingdom, [s]University of Melbourne, Victoria 3010, Australia, [t]Muons, Inc., Batavia, IL 60510, USA, [u]Nagasaki Institute of Applied Science, Nagasaki 851-0193, Japan, [v]National Research Nuclear University, Moscow 115409, Russia, [w]Northwestern University, Evanston, IL 60208, USA, [x]University of Notre Dame, Notre Dame, IN 46556, USA, [y]Universidad de Oviedo, E-33007 Oviedo, Spain, [z]CNRS-IN2P3, Paris, F-75205 France, [aa]Universidad Tecnica Federico Santa Maria, 110v Valparaiso, Chile, [bb]Sejong University, Seoul, 143-747 South Korea, [cc]The University of Jordan, Amman 11942, Jordan, [dd]Universite catholique de Louvain, 1348 Louvain-La-Neuve, Belgium, [ee]University of Zürich, 8006 Zürich, Switzerland, [ff] Massachusetts General Hospital, Boston, MA 02114 USA, [gg]Harvard Medical School, Boston, MA 02114 USA, [hh]Hampton University, Hampton, VA 23668, USA, [ii]Los Alamos National Laboratory, Los Alamos, NM 87544, USA, [jj]Università degli Studi di Napoli Federico I, I-80138 Napoli, Italy

**Abstract**

We study charged particle production ($p_T > 0.5$ GeV/c, $|\eta| < 0.8$) in proton-antiproton collisions at 300 GeV, 900 GeV, and 1.96 TeV. We use the direction of the charged particle with the largest transverse momentum in each event to define three regions of $\eta$-$\phi$ space; "toward", "away", and "transverse". The average number and the average scalar $p_T$ sum of charged particles in the transverse region are sensitive to the modeling of the "underlying event". The transverse region is divided into a MAX and MIN transverse region, which helps separate the "hard component" (initial and final-state radiation) from the "beam-beam remnant" and multiple parton interaction components of the scattering. The center-of-mass energy dependence of the various components of the event are studied in detail. The data presented here can be used to constrain and improve QCD Monte Carlo models, resulting in more precise predictions at the LHC energies of 13 and 14 TeV.

# I. Introduction

The total antiproton-proton cross section is the sum of the elastic and inelastic components, $\sigma_{tot} = \sigma_{EL} + \sigma_{IN}$. Three distinct processes contribute to the inelastic cross section: single diffraction, double-diffraction, and everything else (referred to as "non-diffractive"). For elastic scattering neither of the beam particles break apart (*i.e.,* color singlet exchange). For single and double diffraction one or both of the beam particles are excited into a high-mass color-singlet state (*i.e.,* N* states) which then decay. Single and double diffraction also correspond to color singlet exchange between the beam hadrons. When color is exchanged, the outgoing remnants are no longer color singlets and one has a separation of color resulting in a multitude of quark-antiquark pairs being produced out of the vacuum. The non-diffractive component, $\sigma_{ND}$, involves color exchange and the separation of color, and has both a soft and hard component. Most of the time the color exchange between partons in the beam hadrons occurs through a soft interaction (*i.e.,* no high transverse momentum) and the two beam hadrons move through each other producing soft particles with a uniform distribution in rapidity together with many particles at small angles to the beam. Occasionally, there is a hard scattering among the constituent partons producing outgoing particles and "jets" with high transverse momentum.

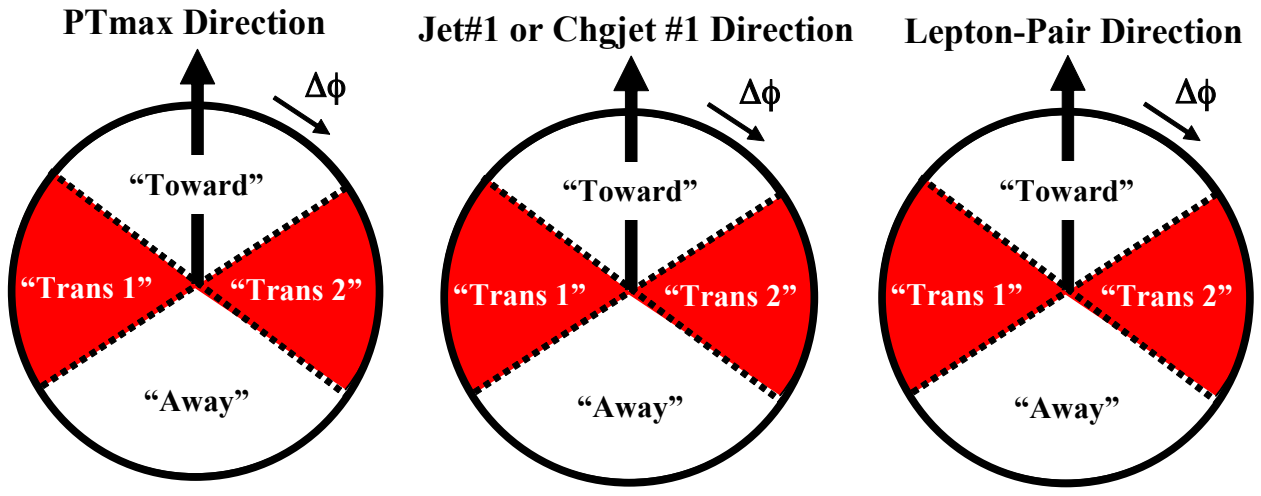

**Figure 1:** Illustration of the regions of η-φ space that are defined relative to the direction of a "leading object" in the event. The "leading object" can be the highest $p_T$ charged particle (*left*), the highest $p_T$ charged-particle or calorimeter jet (*middle*), or the lepton-pair in Z-boson production (*right*). The relative azimuthal angle $\Delta\phi = \phi - \phi_L$, where $\phi_L$ is the azimuthal angle of the leading object and $\phi$ is the azimuthal angle of a charged particle. The "toward" region is defined by $|\Delta\phi| < 60^o$ and $|\eta| < \eta_{cut}$, while the "away" region is $|\Delta\phi| > 120^o$ and $|\eta| < \eta_{cut}$. The two "transverse" regions $-120^o < \Delta\phi < -60^o$, $|\eta| < \eta_{cut}$ and $60^o < \Delta\phi < 120^o$, $|\eta| < \eta_{cut}$ are referred to as "transverse 1" and "transverse 2".

Minimum bias (MB) is a generic term which refers to events that are collected with an online event selection that accepts a large fraction of the overall inelastic cross section with minimal distortion of the general features of the collision. The Collider Detector at Fermilab (CDF) MB online event selection (*i.e.,* trigger) requires at least one charged particle in the forward region $3.2 < \eta < 5.9$ and simultaneously at least one charged particle in the backward region $-5.9 < \eta < -3.2$, where the pseudorapidity $\eta = -\log(\tan(\theta_{cm}/2))$ and $\theta_{cm}$ is the center-of-mass scattering angle. The underlying event (UE) consists of the beam-beam remnants (BBR) and the multiple parton interactions (MPI) that accompany a hard scattering [1]. The UE is an unavoidable background to hard-scattering collider events. To study the UE we use MB data, however, MB and UE observables receive contributions from quite different sources. The majority of MB collisions are soft, while the UE is studied in events in which a hard scattering has occurred. One uses the topological structure of the hard hadron-hadron collision to study the UE experimentally. As illustrated in Fig. 1, on an event-by-event basis, a "leading object" is used to define regions of η-φ space, where η is the pseudorapidity and φ is the azimuthal scattering angle. In Run 1 at CDF we looked at charged particles and used the highest-transverse-momentum charged-particle jet as the leading object [2]. Later in Run 2 we studied the UE using the highest-transverse-energy calorimeter jet as the leading object, and also used the lepton-pair in Z-boson production for the leading object [3]. Here we study charged particles and, as shown in Fig. 2, we use the highest transverse momentum charged particle in the event as the leading object.

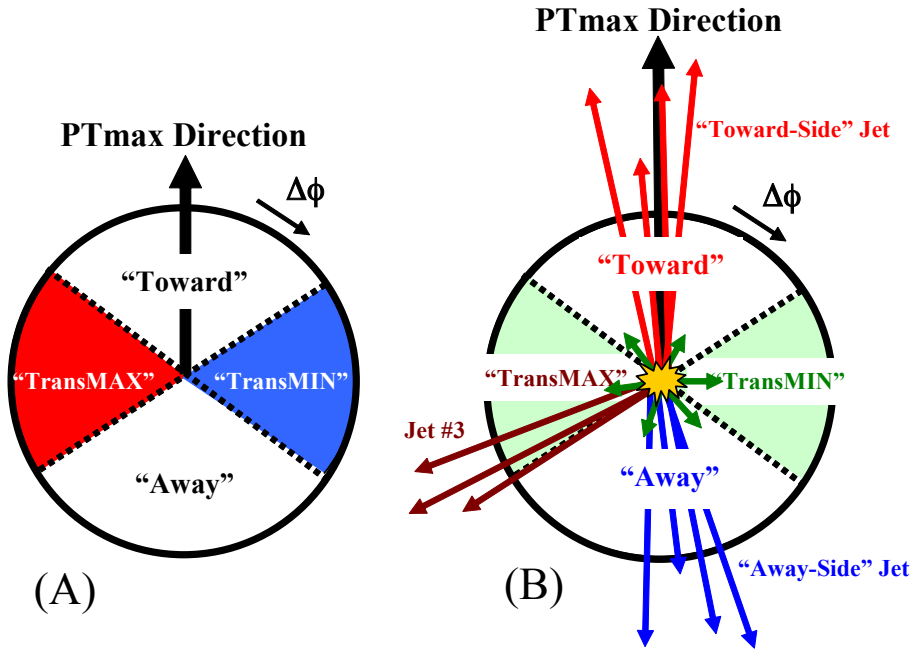

**Figure 2.** **(A)** Illustration of the regions of η-ϕ space that are defined relative to the direction of the highest $p_T$ charged particle (*i.e.,* leading charged particle). The relative angle $\Delta\phi = \phi - \phi_{MAX}$, where $\phi_{MAX}$ is the azimuthal angle of the leading charged particle and ϕ is the azimuthal angle of a charged particle. On an event by event basis, we define "transMAX" ("transMIN") to be the maximum (minimum) number or scalar $p_T$ sum of charged particles in the two transverse regions "transverse 1" and "transverse 2" shown in Fig. 1. **(B)** Illustration of the topology of a hadron-hadron collision in which a hard parton-parton collision has occurred. For events with large initial or final-state radiation the transMAX region contains the third jet, while both the transMAX and transMIN regions receive contributions from the multiple parton interactions (MPI) and the beam-beam remnants (BBR).

The MB and UE observables that we study in this analysis are defined in Table 1. We look at charged particles with $p_T > 0.5$ GeV/c and $|\eta| < \eta_{cut}$. The CDF detector can measure charged particles in the region $|\eta| < 1.1$, however, in order to compare directly with LHC UE data in this analysis we restrict ourselves to $\eta_{cut} = 0.8$. Furthermore, the events considered are required to contain at least one charged particle with $p_T > 0.5$ GeV/c and $|\eta| < 0.8$. We begin by looking at the average overall total number of charged particles and the pseudorapidity distribution of the charged particles. We then examine how the average overall total number of charged particles depends on the center-of-mass energy and on the transverse momentum of the leading charged particle, PTmax. Then, we study the "associated" charged particle and charged PTsum densities, where PTsum is the scalar $p_T$ sum of the charged particles. Densities are formed by dividing by the corresponding area in η-ϕ space. For the overall associated density the area is $\Delta\eta\Delta\phi = 2\eta_{cut}\times 2\pi$. The leading charged particle is not included in the associated density. The associated density is a measure of the number and PTsum of charged particles accompanying (but not including) the highest transverse momentum charged particle.

As shown in Fig. 1, the overall associated density is divided into the "toward", "away", and "transverse" densities. In constructing the transverse density one adds together the two transverse regions: "transverse 1" ($-120^o < \Delta\phi < -60^o$, $|\eta| < \eta_{cut}$) and "transverse 2" ($60^o < \Delta\phi < 120^o$, $|\eta| < \eta_{cut}$). Each of the three regions, toward, away, and transverse have an area of $\Delta\eta\Delta\phi = 2\eta_{cut}\times 2\pi/3$. By comparing these three regions we learn about the topology of the hard-scattering event. As PTmax increases the toward and away densities become much larger than the transverse density since, on average, they receive significant contributions from the two, leading, hard-scattered jets. The toward region contains the toward-side jet, while the away region contains the away-side jet. The number and PTsum densities of charged particles in the transverse region are sensitive to the modeling of the UE.

The transverse region is further separated into the "transMAX" and "transMIN" regions. As shown in Fig. 2, on an event by event basis, we define transMAX (transMIN) to be the transverse region (1 or 2) having the maximum (minimum) of either the number of charged particles, or scalar $p_T$ sum of charged particles, depending on the quantity under study. Again densities are formed by dividing by the area in η-ϕ space, where the transMAX and transMIN regions each have an area of $\Delta\eta\Delta\phi = 2\eta_{cut}\times 2\pi/6$. Hence, the transverse density (also referred to as "transAVE") is the average of the transMAX and the transMIN densities. For events with large initial or final-state radiation the transMAX region often contains the third jet, while both the transMAX and transMIN regions receive contributions from the MPI and BBR components. Thus, the observables in the transMIN region are more sensitive to the MPI and BBR components of the UE, while the "transDIF" observables (transMAX minus the transMIN) are more sensitive to initial-state radiation (ISR) and final-state radiation (FSR) [4].

**Table 1**. Description of the observables studied in this analysis.

| Observable | Description |
|---|---|
| Nchg | Overall number of charged particles ($p_T > 0.5$ GeV/c, $\lvert\eta\rvert < \eta_{cut}$) for events with at least one charged particle ($p_T > 0.5$ GeV/c, $\lvert\eta\rvert < \eta_{cut}$) |
| dN/dη | Number of charged particles ($p_T > 0.5$ GeV/c, $\lvert\eta\rvert < \eta_{cut}$) per unit $\eta$ for events with at least one charged particle ($p_T > 0.5$ GeV/c, $\lvert\eta\rvert < \eta_{cut}$) |
| Overall Associated NchgDen & PTsumDen | Number of charged particles and the scalar $p_T$ sum of charged particles per unit $\eta$-$\phi$ ($p_T > PT_{cut}$, $\lvert\eta\rvert < \eta_{cut}$) that accompany the leading charged particle (*excluding the leading charged particle*) |
| Toward NchgDen & PTsumDen | Number of charged particles and the scalar $p_T$ sum of charged particles per unit $\eta$-$\phi$ in the toward region ($p_T > PT_{cut}$, $\lvert\eta\rvert < \eta_{cut}$) as defined by the leading charged particle (*excluding the leading charged particle*) |
| Away NchgDen & PTsumDen | Number of charged particles and the scalar $p_T$ sum of charged particles per unit $\eta$-$\phi$ in the away region ($p_T > PT_{cut}$, $\lvert\eta\rvert < \eta_{cut}$) as defined by the leading charged particle |
| TransAVE NchgDen & PTsumDen | Number of charged particles and the scalar $p_T$ sum of charged particles per unit $\eta$-$\phi$ in the transverse region ($p_T > PT_{cut}$, $\lvert\eta\rvert < \eta_{cut}$) as defined by the leading charged particle |
| TransMAX NchgDen & PTsumDen | Number of charged particles and the scalar $p_T$ sum of charged particles per unit $\eta$-$\phi$ in the transMAX region ($p_T > PT_{cut}$, $\lvert\eta\rvert < \eta_{cut}$) as defined by the leading charged particle |
| TransMIN NchgDen & PTsumDen | Number of charged particles and the scalar $p_T$ sum of charged particles per unit $\eta$-$\phi$ in the transMIN region ($p_T > PT_{cut}$, $\lvert\eta\rvert < \eta_{cut}$) as defined by the leading charged particle |
| TransDIF NchgDen & PTsumDen | Difference between the number of charged particles and the scalar $p_T$ sum of charged particles per unit $\eta$-$\phi$ in the transMAX and transMIN regions (transDIF = transMAX - transMIN) |
| Transverse $\langle p_T \rangle$ | Average $p_T$ of charged particles in the transverse region ($p_T > PT_{cut}$, $\lvert\eta\rvert < \eta_{cut}$). Require at least 1 charged particle |

QCD Monte Carlo generators such as PYTHIA [5] have parameters which may be adjusted to control the behavior of their event modeling. A specified set of these parameters that has been adjusted to better fit some aspects of the data is referred to as a "tune" [6-8]. The CDF PYTHIA 6.2 Tune A was determined by fitting the CDF Run 1 UE data [2] and the PYTHIA 6.2 Tune DW does a good job in describing both the CDF Run 1 and Run 2 UE data [3]. However, Tune DW does not reproduce perfectly all the features of the LHC data. After the LHC data became available, improved LHC UE tunes were constructed [9, 10]. Tune Z1 and Tune Z2* are PYTHIA 6.4 tunes that were constructed by fitting CMS UE data at 900 GeV and 7 TeV [11]. Tune Z1 uses the CTEQ5L [12] parton distributions (PDFs), while Tune Z2* uses CTEQ6L.

Tune 4C* (CTEQ6L) is a PYTHIA 8 [13] tune which was also determined by fitting CMS UE data at 900 GeV and 7 TeV. The UE observables depend on the PDFs. If one changes the PDFs then one must change the tune. Tune 4C* is similar to Tune 4C [14], but does a slightly better job fitting the CMS UE data at 900 GeV. It takes two center-of-mass energies to determine the energy-dependent MPI parameters of the QCD Monte Carlo models and at least three center-of-mass energies to test the energy dependence of the models. The data presented here can be used to constrain and improve the QCD Monte Carlo models, resulting in more precise predictions at the LHC energies of 13 and 14 TeV.

In Section II we discuss the details of the analysis and explain how the data are corrected to the stable-particle level and how the systematic errors are determined. The analysis techniques employed here are similar to those used in our previous CDF Run 2 UE analysis [3]. The data and comparisons with the PYTHIA tunes are shown in Section III. Section IV contains a summary and conclusions.

# II. ANALYSIS DETAILS

## (1) Data and Vertex Selection

The CDF Run 2 detector became operational in 2001. It is an azimuthally and forward-backward-symmetric solenoidal particle detector [15] combining precision charged-particle tracking with fast projective calorimetry and fine grained muon detection. Tracking systems are designed to detect charged particles and measure their momenta and displacements from the point of collision, termed the primary interaction vertex. The tracking system consists of a silicon microstrip system (not used for this analysis) and an open-cell wire drift chamber, the latter called the Central Outer Tracker (COT) that surrounds the silicon. The positive z-axis is defined to lie along the incident proton beam direction. We use all the 300 GeV and 900 GeV MB data resulting from a dedicated data-taking period in which the collider was operated at reduced energy (referred to as the "Tevatron Energy Scan"). At 1.96 TeV we include the 2 $fb^{-1}$ of Run 2 MB data that was taken before January 30, 2007, where the instantaneous luminosity was not large so that the pile-up corrections are small (see Sec. II.3). In order to estimate the systematic uncertainties, at each of the three energies we consider two different vertex selection criteria. One selection requires zero or one high-quality vertices within the fiducial region $|Z_{vertex}| \leq 60$ cm centered around the nominal CDF z = 0. The other selection requires events to have one and only one high-quality vertex within $|Z_{vertex}| \leq 60$ cm.

## (2) Track-Selection Criteria (Loose and Tight)

We consider charged tracks that have been measured by the central outer tracker (COT). The COT [16] is a cylindrical open-cell drift chamber with 96 sense wire layers grouped into eight alternating superlayers of stereo and axial wires. Its active volume covers $40 < r < 137$ cm, where r is the radial coordinate in the plane transverse to the z axis, and $|z| < 155$ cm, thus providing fiducial coverage in $|\eta| \leq 1.1$ to tracks originating within $|z| \leq 60$ cm. We include tracks in the region $0.5 < p_T < 150$ GeV/c and $|\eta| < 0.8$, where the COT has high efficiency. At higher values of $p_T$ the track momentum resolution deteriorates. The upper limit of 150 GeV/c is chosen to prevent mis-measured tracks with very high $p_T$ from distorting the average charged-particle density and the average charged-particle PTsum density. Tracks are required to be reconstructed with COT signals from at least 10 axial wires from two axial segments and 10 stereo wires from 2 stereo segments. In addition, the tracks are required to point back to the primary vertex in the event. In order to estimate the systematic uncertainties, we employ both a "loose" and a "tight" track selection criterion. The loose track selection requires $|d_0| < 1.0$ cm and $|z - z_{vertex}| < \Delta Z_{cut} = 3.0$ cm, where $d_0$ is the beam corrected transverse impact parameter and $z - z_{vertex}$ is the distance on the z-axis (beam axis) between the track projection and the primary vertex. The tight track selection requires that $|d_0| < 0.5$ cm and $|z - z_{vertex}| < \Delta Z_{cut} = 2.0$ cm. This is identical to our previous Run 2 UE analysis [3]. For both the tight and loose cases the transverse impact parameter is corrected for the beam position. For events with no high-quality vertex we require $|z - z_{max}| < 2\Delta Z_{cut}$, where $z - z_{max}$ is the longitudinal distance between the measured track and the highest-$p_T$ track (*i.e.,* leading track).

Three data sets are considered in this analysis at each of the three energies: 1.96 TeV, 900 GeV, and 300 GeV. The first requires 0 or 1 high-quality vertices and uses the tight track selection criterion (data set T01). The second also requires 0 or 1 high-quality vertices, but uses loose track selection criterion (data set L01). The third requires 1 and only 1 high-quality vertex and uses tight track selection criterion (data set T1). Requiring at least one high quality vertex biases the data toward more active events. Most events with large PTmax have at least one high quality vertex and hence the data sets T01 and T1 become the same for PTmax > 4 GeV/c. The data sets T01 and L01 differ slightly at all PTmax values. The loose track selection criterion accept slightly more tracks than the tight track selection criterion. The T01 data set is the primary data of this analysis. The L01 and T1 data sets are used to evaluate systematic errors, as discussed in Sec. II (5).

## (3) Pile-Up Corrections at 1.96 TeV

Although we require zero or one high-quality reconstructed vertex, the observables in Table 1 are still affected by pile-up (*i.e.,* more that one proton-antiproton collision in the event). Tracks are required to point back to the primary vertex, but the track observables are affected by pile-up when two vertices overlap. Vertices within about 3.0 cm of each other merge together as one. Large instantaneous luminosity implies more pile-up. The data in each PTmax bin are plotted versus the instantaneous luminosity and fit to a straight line. This function is then used to correct the data for pile-up on an event-by-event basis. The value of every bin of the plots at 1.96 TeV have been corrected for pile-up. In all cases the pile-up corrections are less than 4%. The instantaneous luminosities at 300 GeV and 900 GeV are so small that there is no need for pile-up corrections of the data.

## (4) Correcting to the Particle Level (Response and Correction Factors)

The charged tracks measured in the CDF detector are corrected to the stable-particle level using the same bin-by-bin method we used in our previous Run 2 UE analysis [3]. We rely on PYTHIA Tune A and the CDF detector simulation CDFSIM (parameterized response of the CDF II detector [17, 18]) to correct the measured tracks back to the prompt stable charged particle level. Particles are considered stable if $c\tau > 10$ mm (*i.e.,* $K_s$, $\Lambda$, $\Sigma$, $\Xi$, and $\Omega$ are considered stable). PYTHIA Tune A is used to calculate the observables in Table 1 at the particle level in bins of the highest-$p_T$ charged particle (GEN) and at the detector level in bins of the highest-$p_T$ track (CDFSIM). The detector-level data in bins of the highest-$p_T$ track are corrected by multiplying by the correction factor, GEN/CDFSIM. Smooth curves are drawn through the QCD

Monte Carlo predictions at both the generator level (GEN) and the detector level (CDFSIM) to aid in comparing the theory with the data and also to construct the correction factors. Correction factors for every bin of every observable in Table 1 are constructed for each of the three data sets (T01, L01, and T1) at the three center-of-mass energies: 1.96 TeV, 900 GeV, and 300 GeV. At 1.96 TeV correction factors are constructed after correcting for pile-up. The correction factors depend on the $p_T$ of the leading charged particle, PTmax. For the T01 and L01 data sets the corrections are less than 10% for all values of PTmax. For PTmax > 2 GeV/c the corrections to the T1 data set are less than 10%, but at low PTmax values the corrections are around 20%. The data presented here correspond to the corrected T01 data set. The corrected T1 and L01 data sets are used to estimate the systematic uncertainties. The data points are plotted at the center of the bins.

### (5) Systematic Uncertainties

The three datasets (T01, L01, and T1) are each corrected to the particle level using their corresponding correction factors. If PYTHIA Tune A fit the data perfectly and if the detector simulation (CDFSIM) were perfect, then the corrected data from the three data sets would be identical. The differences among the three corrected datasets are used to estimate the systematic uncertainties. The first systematic uncertainty (sys1) is a measure of how well CDFSIM simulates the difference between the loose and tight track selection (bin-by-bin difference between the corrected data sets L01 and T01). The second (sys2) is a measure of how well CDFSIM simulates the difference in including or excluding events with zero high-quality vertices (bin-by-bin difference between the corrected data sets T1 and T01). The third (sys3 = 2%) is included to take into account the accuracy of constructing the smooth theory curves that are used to construct the response and correction factors. The overall total uncertainty results from adding the statistical error in quadrature with the three systematic uncertainties: sys1, sys2, and sys3. At low PTmax values the overall error is dominated by sys2, while at large PTmax the overall error is predominately statistical.

## III. Results and Comparisons

### (1) Total Number of Charged Particles

Figure 3 shows the data at 1.96 TeV, 900 GeV, and 300 GeV on the pseudorapidity distribution, $dN/d\eta$, for charged particles with $|\eta| < 0.8$ and $p_T > 0.5$ GeV/c and $p_T > 1.0$ GeV/c for events with at least one charged particle with $|\eta| < 0.8$ and $p_T > 0.5$ GeV/c and $p_T > 1.0$ GeV/c, respectively, compared with PYTHIA Tune Z1 [19]. The pseudorapidity distribution is shown for both $p_T > 0.5$ GeV/c and $p_T > 1.0$ GeV/c in order to test if the models give the correct transverse-momentum distribution of the charged particles, and as can be seen in Fig. 3, the data have a slightly steeper $p_T$ distribution than does Tune Z1. The data on the pseudorapidity distribution, $dN/d\eta$, at $\eta = 0$ plotted versus the center-of-mass energy are also shown. The pseudorapidity distribution increases slowly with energy and PYTHIA Tune Z1 describes the rise with energy fairly well. The $dN/d\eta$ distributions correspond to the average number of charged particles per unit $\eta$ and are normalized so that the integral is equal to the overall average number of charged particles with $|\eta| < 0.8$ and $p_T > 0.5$ GeV/c and with $|\eta| < 0.8$ and $p_T > 1.0$ GeV/c for events with at least one charged particle with $|\eta| < 0.8$ and $p_T > 0.5$ GeV/c and $p_T > 1.0$ GeV/c, respectively, as follows:

$$N_{chg} = \int_{-0.8}^{0.8} \frac{dN}{d\eta} d\eta \qquad (1)$$

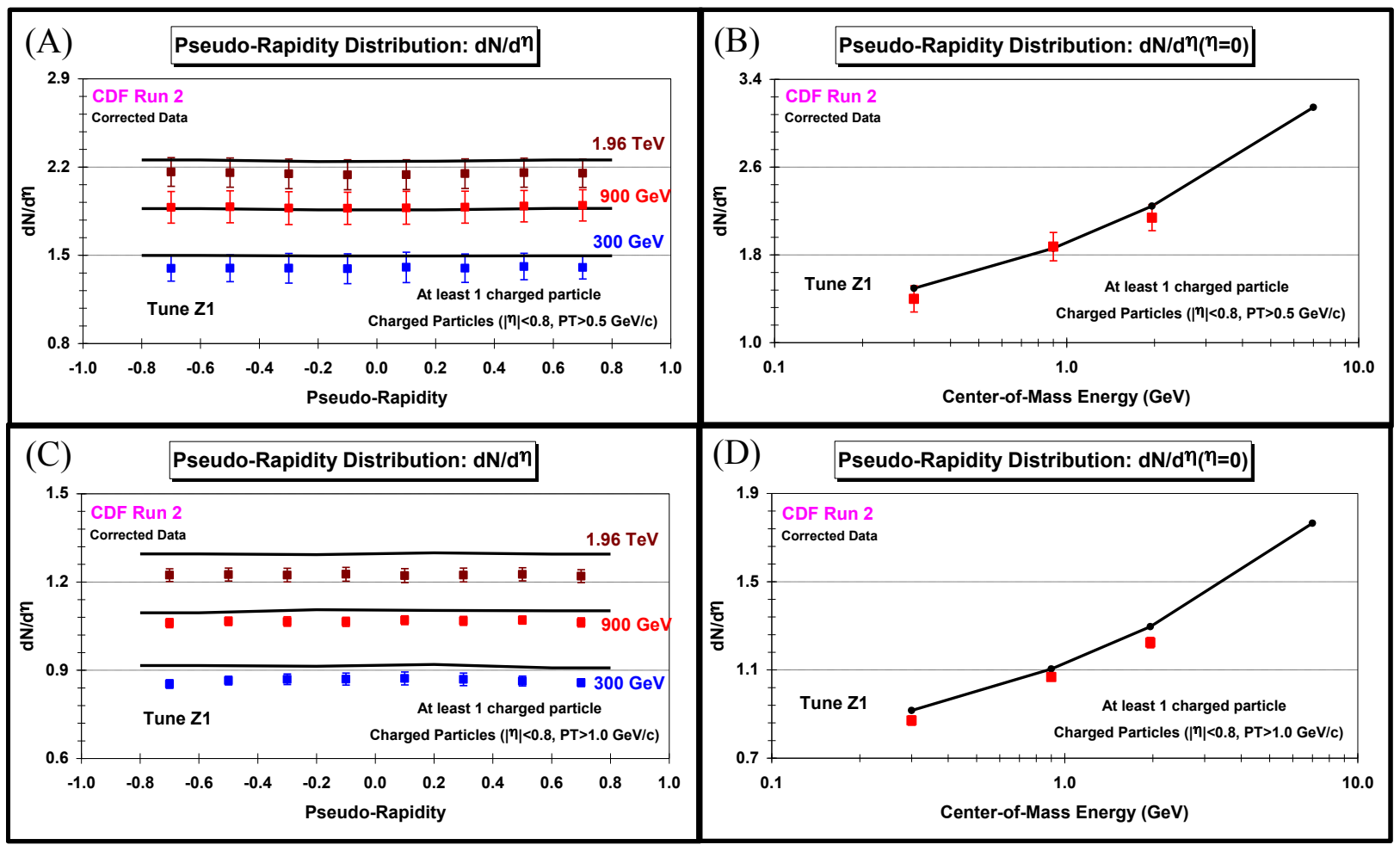

**Figure 3**. **(A)** Data at 1.96 TeV, 900 GeV, and 300 GeV on the pseudorapidity distribution, dN/dη, for charged particles with $|\eta| < 0.8$ and $p_T > 0.5$ GeV/c and **(C)** $p_T > 1.0$ GeV/c for events with at least one charged particle with $|\eta| < 0.8$ and $p_T > 0.5$ GeV/c and $p_T > 1.0$ GeV/c, respectively. **(B)** Data on the pseudorapidity distribution, dN/dη, at $\eta = 0$ for charged particles with $|\eta| < 0.8$ and $p_T > 0.5$ GeV/c and **(D)** $p_T > 1.0$ GeV/c for events with at least one charged particle with $|\eta| < 0.8$ and $p_T > 0.5$ GeV/c and $p_T > 1.0$ GeV/c, respectively, plotted versus the center-of-mass energy. The data are corrected to the particle level with errors that include both the statistical error and the systematic uncertainty and are compared with PYTHIA 6.4 Tune Z1.

In constructing dN/dη we require Nchg ≥ 1 and include all $p_T$ values greater than 0.5 GeV/c of the leading charged particle. This is exactly the same set of charged particles that are included in our study of the UE. To study the UE, however, we look at the number and PTsum of the charged particles in the transverse region as a function of the transverse momentum of the leading charged particle.

**Table 2**. Data at 1.96 TeV, 900 GeV, and 300 GeV on the average overall number of charged particles and the average overall density of charged particle with $|\eta| < 0.8$ and $p_T > 0.5$ GeV/c for events with at least one charged particle with $|\eta| < 0.8$ and $p_T > 0.5$ GeV/c. The data are corrected to the particle level with errors that include both the statistical error and the systematic uncertainty.

| Ecm | Nchg | NchgDen |
|---|---|---|
| 300 GeV | 2.24 ± 0.16 | 0.22 ± 0.02 |
| 900 GeV | 3.01 ± 0.20 | 0.30 ± 0.02 |
| 1.96 TeV | 3.44 ± 0.19 | 0.34 ± 0.02 |

Table 2 shows the data on the average overall number of charged particles and the average overall density of charged particles with $|\eta| < 0.8$ and $p_T > 0.5$ GeV/c for events with at least one charged particle with $|\eta| < 0.8$ and $p_T > 0.5$ GeV/c. The data are corrected to the particle level with errors that include both the statistical error and the systematic uncertainty. The overall density is computed by dividing by $1.6 \times 2\pi$. The average overall number of charged particles increases by 50% from 2.24 at 300 GeV to 3.44 at 1.96 TeV.

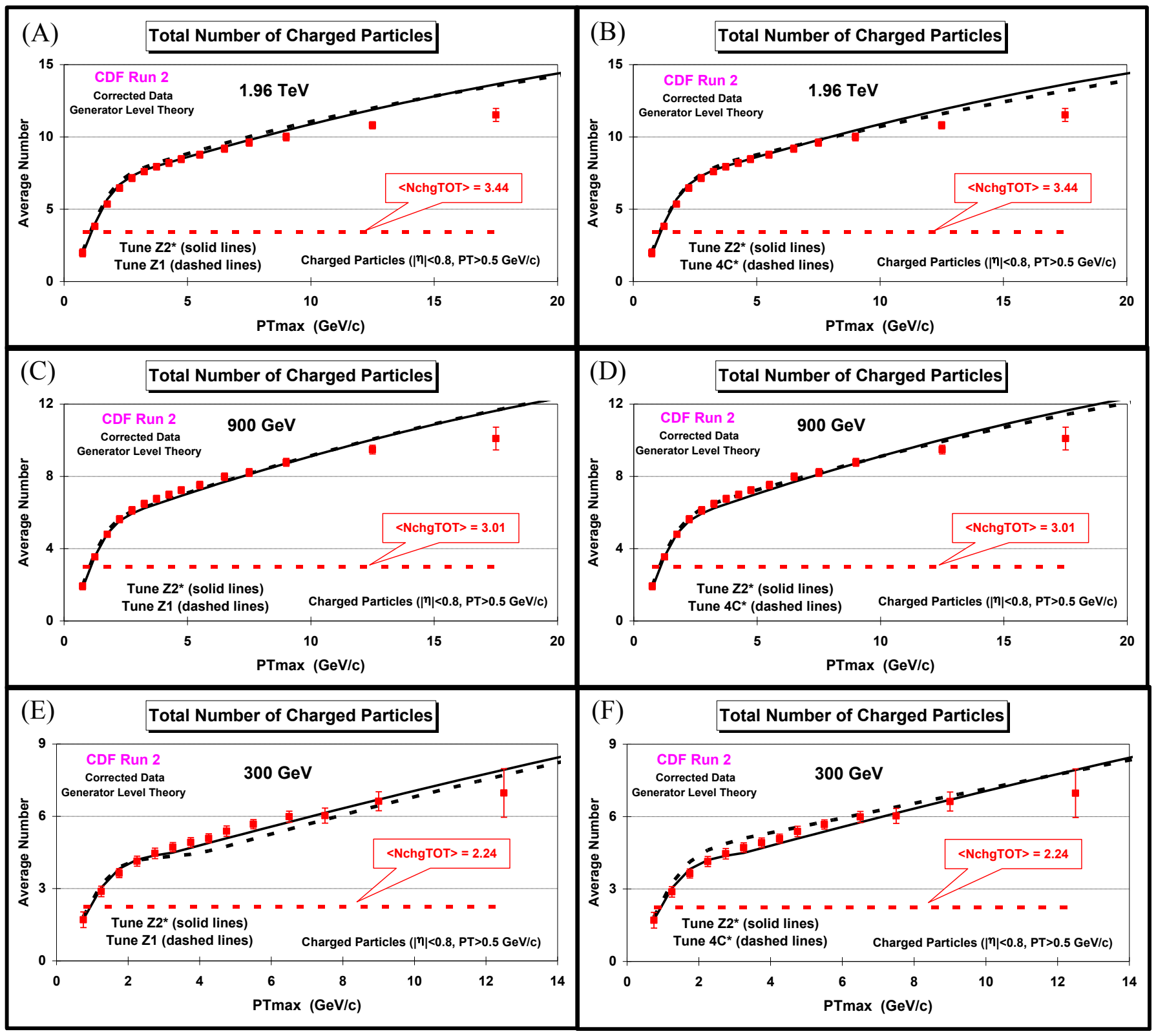

**Figure 4**. Data at 1.96 TeV **(A,B)**, 900 GeV **(C,D)**, and 300 GeV **(E,F)** on the average overall number of charged particles with $|\eta| < 0.8$ and $p_T > 0.5$ GeV/c (including the leading charged particle) for events with at least one charged particle with $|\eta| < 0.8$ and $p_T > 0.5$ GeV/c plotted versus the transverse momentum of the leading charged particle, PTmax. The horizontal dashed lines correspond to the average overall number of charged particles with $|\eta| < 0.8$ and $p_T > 0.5$ GeV/c for events with at least one charged particle with $|\eta| < 0.8$ and $p_T > 0.5$ GeV/c if one includes all PTmax values (see Table 2). The data are corrected to the particle level with errors that include both the statistical error and the systematic uncertainty, and are compared with PYTHIA Tune Z1 and Z2* **(A,C,E)** and PYTHIA Tune Z2* and 4C* **(B,D,F)**.

Figure 4 compares the average overall number of charged particles from Table 2 with the average overall number of charged particles with $|\eta| < 0.8$ and $p_T > 0.5$ GeV/c (including the highest-$p_T$ charged particle) for events with at least one charged particle with $|\eta| < 0.8$ and $p_T >$ 0.5 GeV/c plotted versus the transverse momentum of the leading charged particle, PTmax. The average overall number of charged particles in Table 2 corresponds to including all PTmax values. As one would expect the overall average number of charged particles increases as PTmax increases. For example at 1.96 TeV the overall average number of charged particles is 3.44 if one includes all PTmax values, and events with PTmax $\approx$ 10 GeV/c have, on the average, roughly 10 charged particles. This observable is sensitive to the overall structure of the event. Demanding a hard scattering selects events with higher multiplicity. The QCD Monte Carlo model tunes describe this observable fairly well. However, at 1.96 TeV and 900 GeV the tunes produce slightly too many charged particles at large PTmax values.

Figure 5 shows the data at 1.96 TeV, 900 GeV, and 300 GeV on the overall associated charged particle and charged PTsum densities as defined by the leading charged particle, as a function of the transverse momentum of the leading charged particle, PTmax. The leading charged particle is not included in the overall associated density. This quantity is a measure of the number of particles and PTsum accompanying (but not including) the leading charged particle. The associated charged PTsum density increases more rapidly with increasing PTmax than does the associated charged particle density. This is a reflection of the fact that the average transverse momentum of the charged particles increases as PTmax increases. The QCD Monte Carlo model tunes describe these two observables fairly well. However, at 1.96 TeV and 900 GeV the tunes produce slightly too many associated charged particles at large PTmax values.

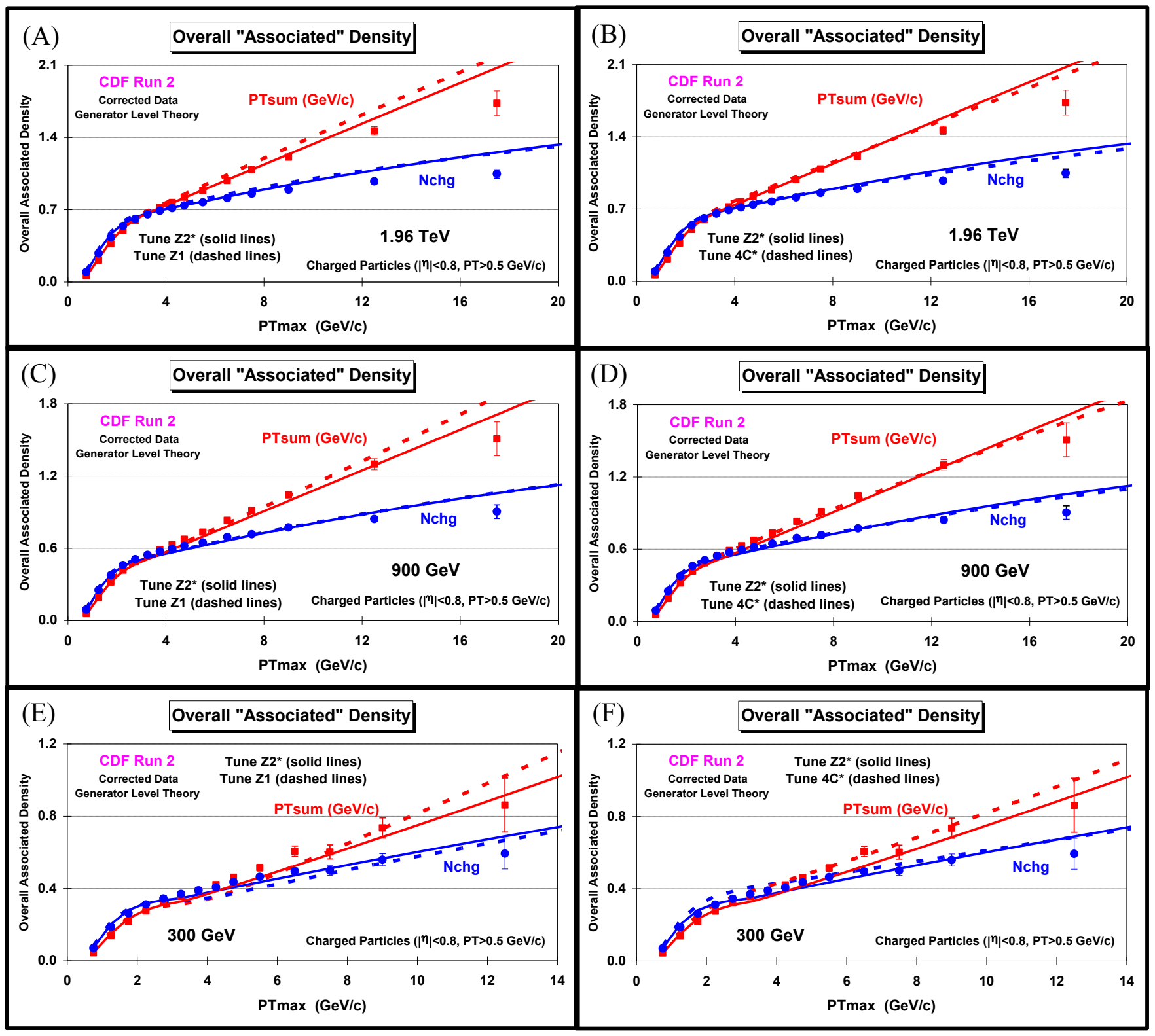

**Figure 5**. Data at 1.96 TeV **(A,B)**, 900 GeV **(C,D)**, and 300 GeV **(E,F)** on the overall associated charged particle and charged PTsum density ($p_T > 0.5$ GeV/c, $|\eta| < 0.8$) as defined by the leading charged particle, as a function of the $p_T$ of the leading charged particle, PTmax (where the vertical axis scale applies to both densities with appropriate units). The leading charged particle is not included in the overall associated density. The data are corrected to the particle level with errors that include both the statistical error and the systematic uncertainty, and are compared with PYTHIA Tune Z1 and Z2* **(A,C,E)** and PYTHIA Tune Z2* and 4C* **(B,D,F)**.

## (2) The Toward and Away Regions

Figures 6 and 7 show the data at 1.96 TeV, 900 GeV, and 300 GeV on the charged particle and the charged PTsum densities in the toward, away, and transverse regions as defined by the leading charged particle, as a function of the transverse momentum of the leading charged particle. The leading charged particle is not included in the toward density. These observables measure the overall topological structure of the event. The toward region contains, on the average, the leading jet in the event, while the away region, on the average, contains the corresponding away-side jet. The transverse (*i.e.,* transAVE) region is perpendicular to the hard-scattering and is sensitive to the UE. The overall associated density in Fig. 5 is the average of the toward, away, and transverse densities.

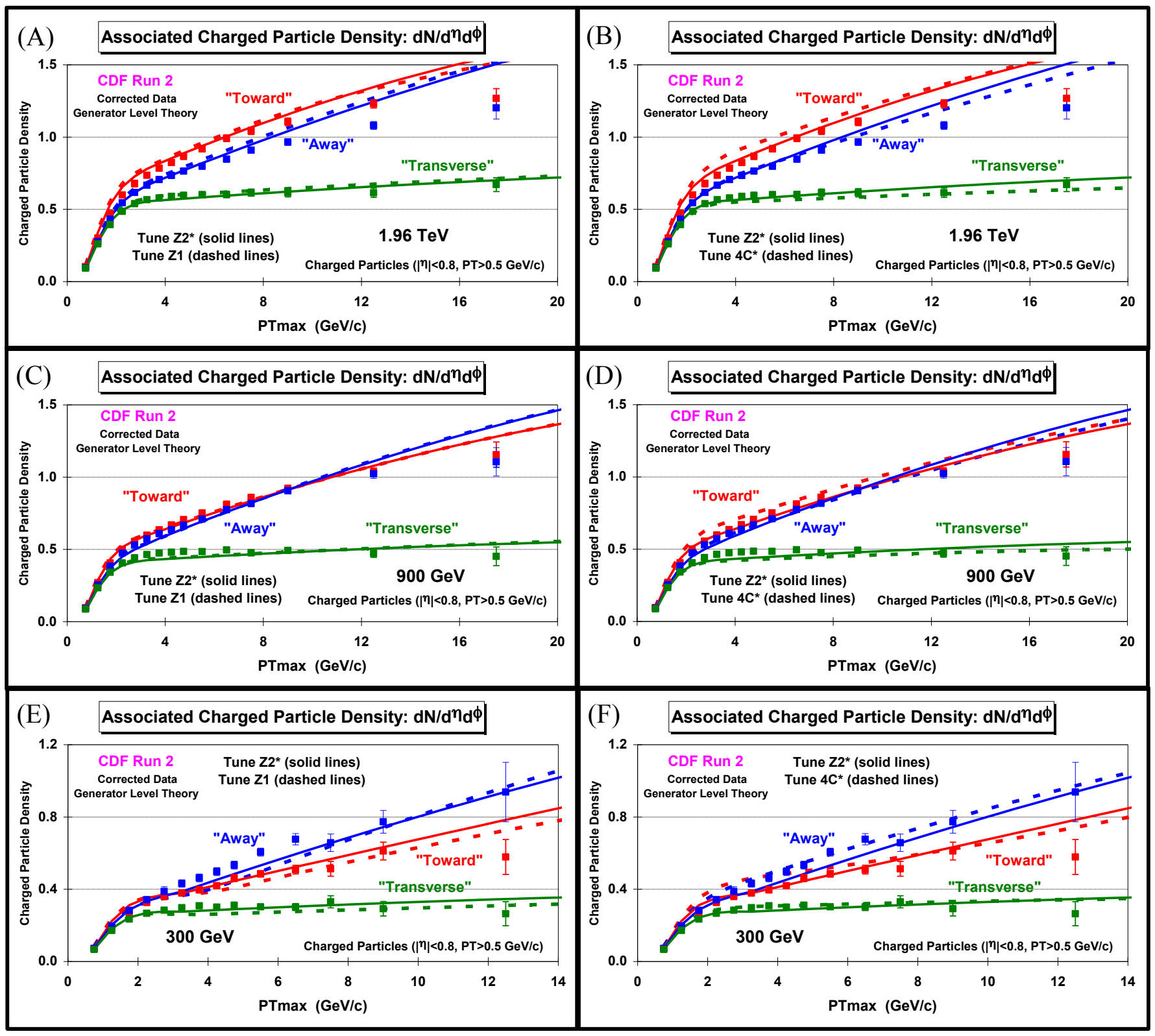

**Figure 6**. Data at 1.96 TeV **(A,B)**, 900 GeV **(C,D)**, and 300 GeV **(E,F)** on the charged particle density ($p_T$ > 0.5 GeV/c, $|\eta| < 0.8$) in the toward, away, and transverse regions as defined by the leading charged particle, as a function of the $p_T$ of the leading charged particle, PTmax. The leading charged particle is not included in the toward density. The data are corrected to the particle level with errors that include both the statistical error and the systematic uncertainty, and are compared with PYTHIA Tune Z1 and Z2* **(A,C,E)** and PYTHIA Tune Z2* and 4C* **(B,D,F)**.

Figures 8 and 9 compare the charged particle density and the charged PTsum density, respectively, in the toward and away regions at the three center-of-mass energies; 1.96 TeV, 900 GeV, and 300 GeV. The charge particle and PTsum densities in the toward region behave differently than they do in the away region, as the center-of-mass energy increases. The UE contributes to the toward and away regions, however, these regions are dominated by hard-scattered jets. The toward region observables measure the number and PTsum of the charged particles accompanying the leading charged particle. The jet in the toward region is not an average jet. It is a jet in which almost all the momentum of the jet is taken by one charged particle. In order to describe this region the QCD Monte Carlo models must describe well the $z \approx 1$ region of the fragmentation function, where z is the fraction of the overall jet momentum carried by a single charged particle. At 300 GeV the PTmax distribution is very steep and the probability of having a leading charged particle with, for example, PTmax ≈ 10 GeV/c is small. The QCD Monte Carlo models describe this by producing a parton with transverse momentum just slightly higher than 10 GeV/c which fragments into a charged particle carrying almost all the momentum of the parton ($z \approx 1$), resulting in very few accompanying jet particles. At 1.96 TeV the PTmax distribution is not as steep and there is a higher probability of having a leading charged particle with PTmax ≈ 10 GeV/c. Here the fraction of the jet momentum carried by the leading charged particle is not as high, and hence there are more accompanying jet particles.

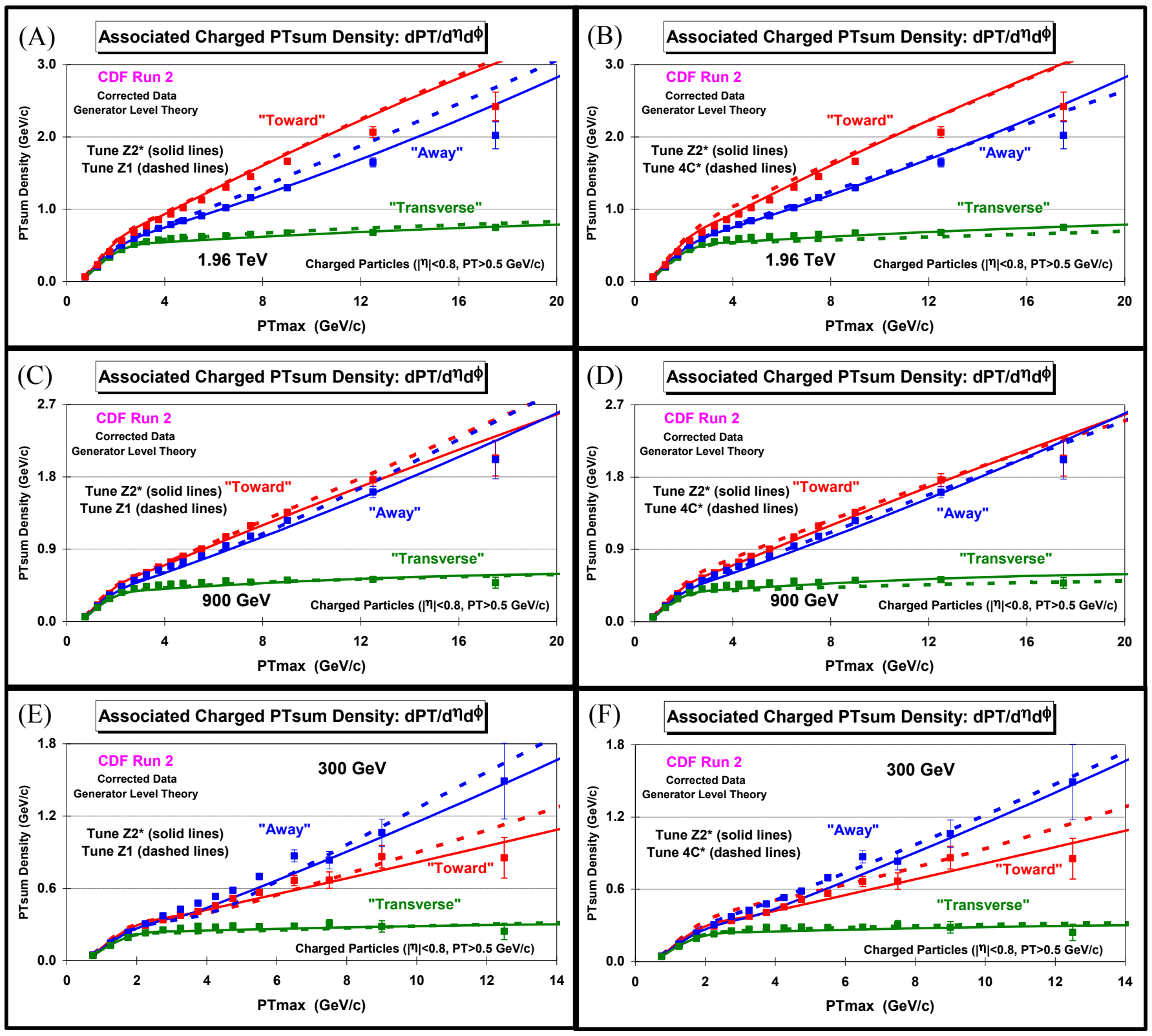

**Figure 7**. Data at 1.96 TeV **(A,B)**, 900 GeV **(C,D)**, and 300 GeV **(E,F)** for the charged PTsum density ($p_T$ > 0.5 GeV/c, $|\eta|$ < 0.8) in the toward, away, and transverse regions as defined by the leading charged particle, as a function of the $p_T$ of the leading charged particle, PTmax. The leading charged particle is not included in the toward density. The data are corrected to the particle level with errors that include both the statistical error and the systematic uncertainty, and are compared with PYTHIA Tune Z1 and Z2* **(A,C,E)** and PYTHIA Tune Z2* and 4C* **(B,D,F)**.

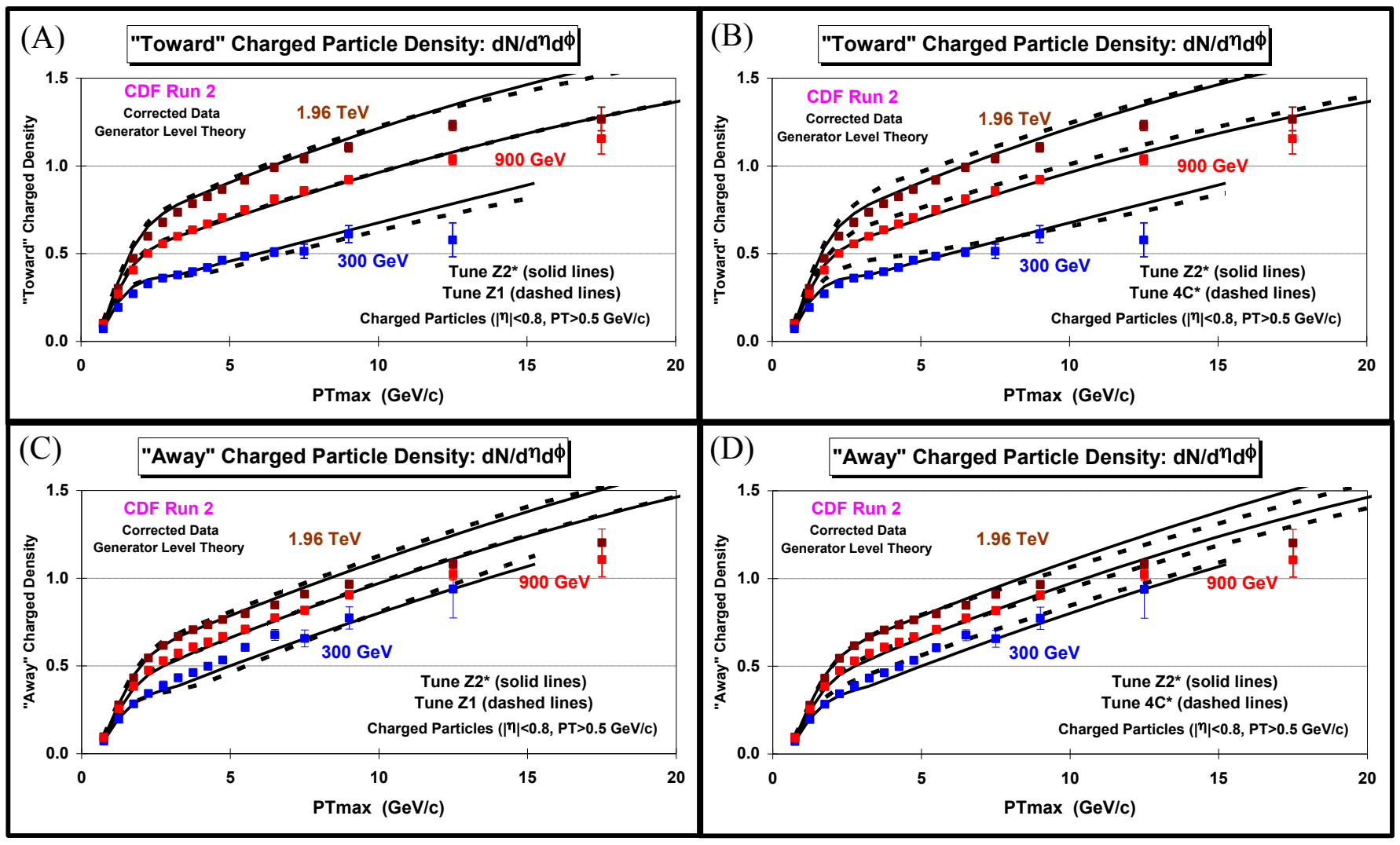

**Figure 8**. Data at 1.96 TeV, 900 GeV, and 300 GeV on the charged particle density ($p_T$ > 0.5 GeV/c, $|\eta|$ < 0.8) in the toward **(A,B)** and away **(C,D)** regions as defined by the leading charged particle, as a function of the $p_T$ of the leading charged particle, PTmax. The leading charged particle is not included in the toward density. The data are corrected to the particle level with errors that include both the statistical error and the systematic uncertainty, and are compared with PYTHIA Tune Z1 and Z2* **(A,C)** and PYTHIA Tune Z2* and 4C* **(B,D)**.

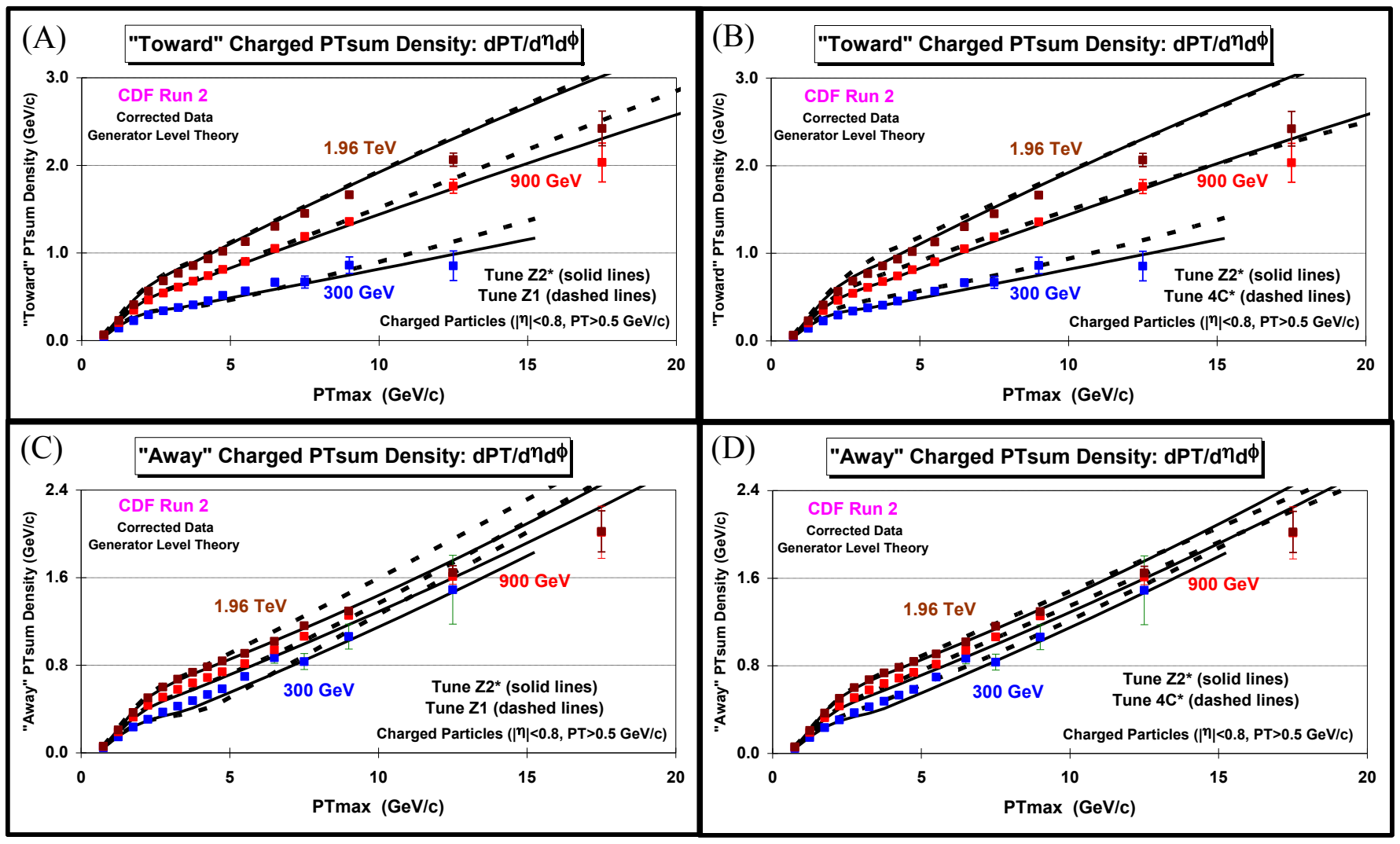

**Figure 9**. Data at 1.96 TeV, 900 GeV, and 300 GeV on the charged PTsum density ($p_T > 0.5$ GeV/c, $|\eta| < 0.8$) in the toward **(A,B)** and away **(C,D)** regions as defined by the leading charged particle, as a function of the $p_T$ of the leading charged particle, PTmax. The leading charged particle is not included in the toward density. The data are corrected to the particle level with errors that include both the statistical error and the systematic uncertainty, and are compared with PYTHIA Tune Z1 and Z2* **(A,C)** and PYTHIA Tune Z2* and 4C* **(B,D)**.

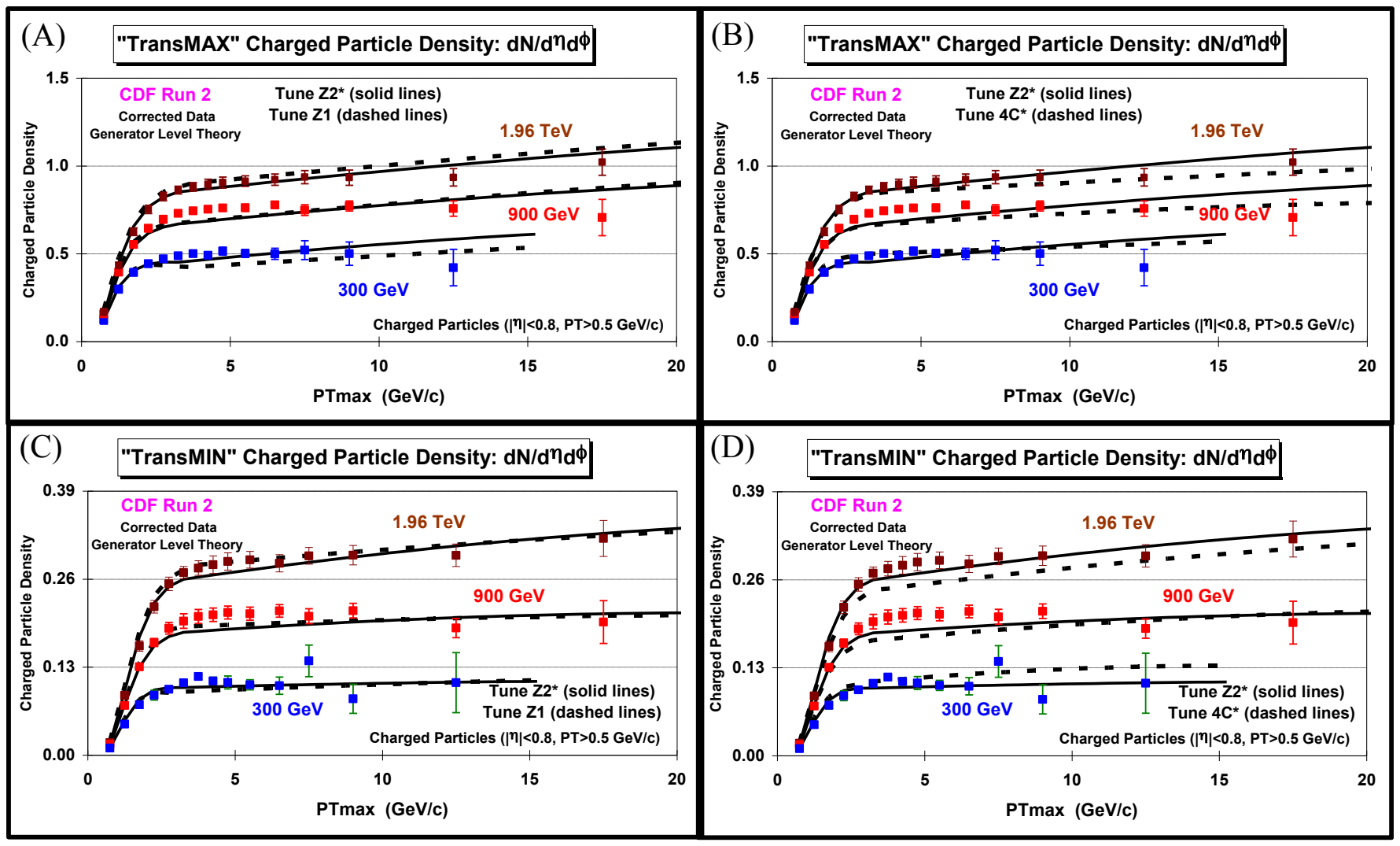

**Figure 10**. Data at 1.96 TeV, 900 GeV, and 300 GeV on the charged particle density ($p_T > 0.5$ GeV/c, $|\eta| < 0.8$) in the transMAX **(A,B)** and transMIN **(C,D)** regions as defined by the leading charged particle, as a function of the $p_T$ of the leading charged particle, PTmax. The data are corrected to the particle level with errors that include both the statistical error and the systematic uncertainty, and are compared with PYTHIA Tune Z1 and Z2* **(A,C)** and PYTHIA Tune Z2* and 4C* **(B,D)**.

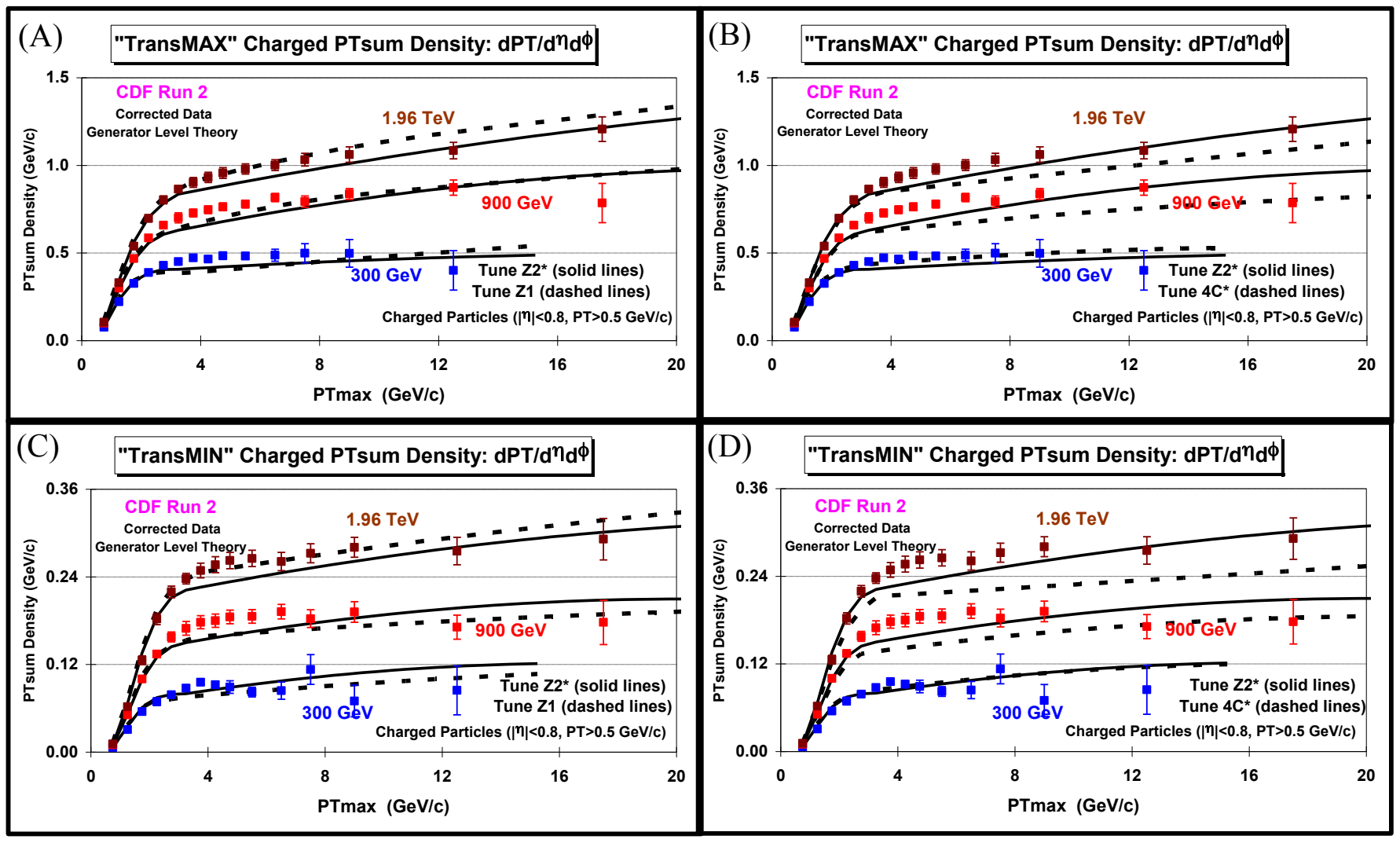

**Figure 11**. Data at 1.96 TeV, 900 GeV, and 300 GeV on the charged PTsum density ($p_T > 0.5$ GeV/c, $|\eta| < 0.8$) in the transMAX **(A,B)** and transMIN **(C,D)** regions as defined by the leading charged particle, as a function of the $p_T$ of the leading charged particle, PTmax. The data are corrected to the particle level with errors that include both the statistical error and the systematic uncertainty, and are compared with PYTHIA Tune Z1 and Z2* **(A,C)** and PYTHIA Tune Z2* and 4C* **(B,D)**.

Unlike the toward-side jet, the away-side jet is an average jet. However, it is not always in the central region $|\eta| < 0.8$. When it is in this region then the away region observables are dominated by the away-side jet. When it is not, then the away region observables are dominated by ISR, FSR, and the UE. The probability that the away-side jet is in the central region is a function of both PTmax and the center-of-mass energy. For PTmax ≈ 10 GeV/c it is more likely that the away-side jet is central at 300 GeV than at 1.96 TeV. At large PTmax values at 300 GeV the charge particle and PTsum densities are larger in the away region than they are in the toward region. At 900 GeV they are roughly the same, and at 1.96 TeV the densities in the toward region are larger than they are in the away region. The QCD Monte Carlo model tunes do a good job in describing the qualitative behavior of the observables in the toward and away regions. There is a tendency for the tunes to produce too much associated density in the toward region.

## (3) transMAX, transMIN, transAVE, and transDIF

Figures 10 and 11 show the data at 1.96 TeV, 900 GeV, and 300 GeV on the charged particle density and charged PTsum density, respectively, in the transMAX and transMIN regions as defined by the leading charged particle, as a function of the $p_T$ of the leading charged particle, PTmax. Figures 12 and 13 show the CDF data at 1.96 TeV, 900 GeV, and 300 GeV on the charged particle density and PTsum density, respectively, for transAVE and transDIF as a function of PTmax. The transAVE density is the average of the transMAX and transMIN densities, while the transDIF density is the transMAX density minus the transMIN density. The transverse density shown in Figs. 6 and 7 corresponds to the transAVE density.

Figures 14 and 15 show data on the transMAX and transMIN charged particle density and charged PTsum density, respectively, as defined by the leading charged particle, for 5.0 < PTmax < 6.0 GeV/c plotted versus the center-of-mass energy. For PTmax < 5.0 GeV/c, the UE observables in the transverse region increase rapidly as PTmax increases, while for PTmax >

5.0 GeV/c they increase slowly with increasing PTmax (*i.e.*, the “plateau” region). The bin 5.0 < PTmax < 6.0 GeV/c is selected since it corresponds to the beginning of the “plateau” region. Figures 16 and 17 show data on the transAVE and transDIF charged particle density and charged PTsum density, respectively, plotted versus the center-of-mass energy. These figures also show the ratio of the data at 1.96 TeV, 900 GeV, and 300 GeV to the corresponding value at 300 GeV. All four densities, MAX, MIN, AVE, and DIF have different center-of-mass energy dependences and the QCD Monte Carlo model tunes do a remarkably good job in describing the general features of these four observables.

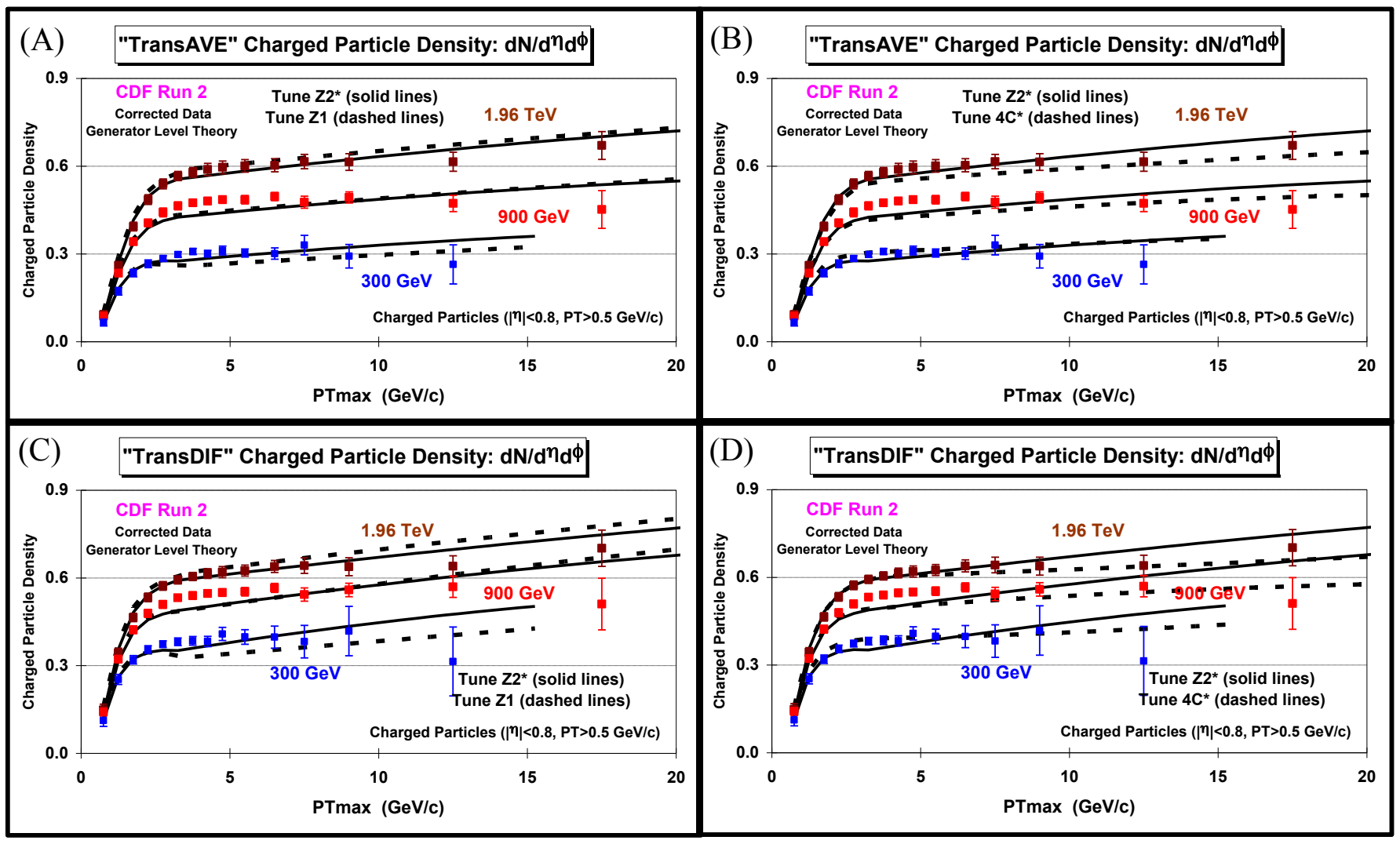

**Figure 12**. Data at 1.96 TeV, 900 GeV, and 300 GeV on the charged particle density ($p_T$ > 0.5 GeV/c, |η| < 0.8) in the transAVE **(A,B)** and transDIF **(C,D)** regions as defined by the leading charged particle, as a function of the $p_T$ of the leading charged particle, PTmax. The data are corrected to the particle level with errors that include both the statistical error and the systematic uncertainty, and are compared with PYTHIA Tune Z1 and Z2* **(A,C)** and PYTHIA Tune Z2* and 4C* **(B,D)**.

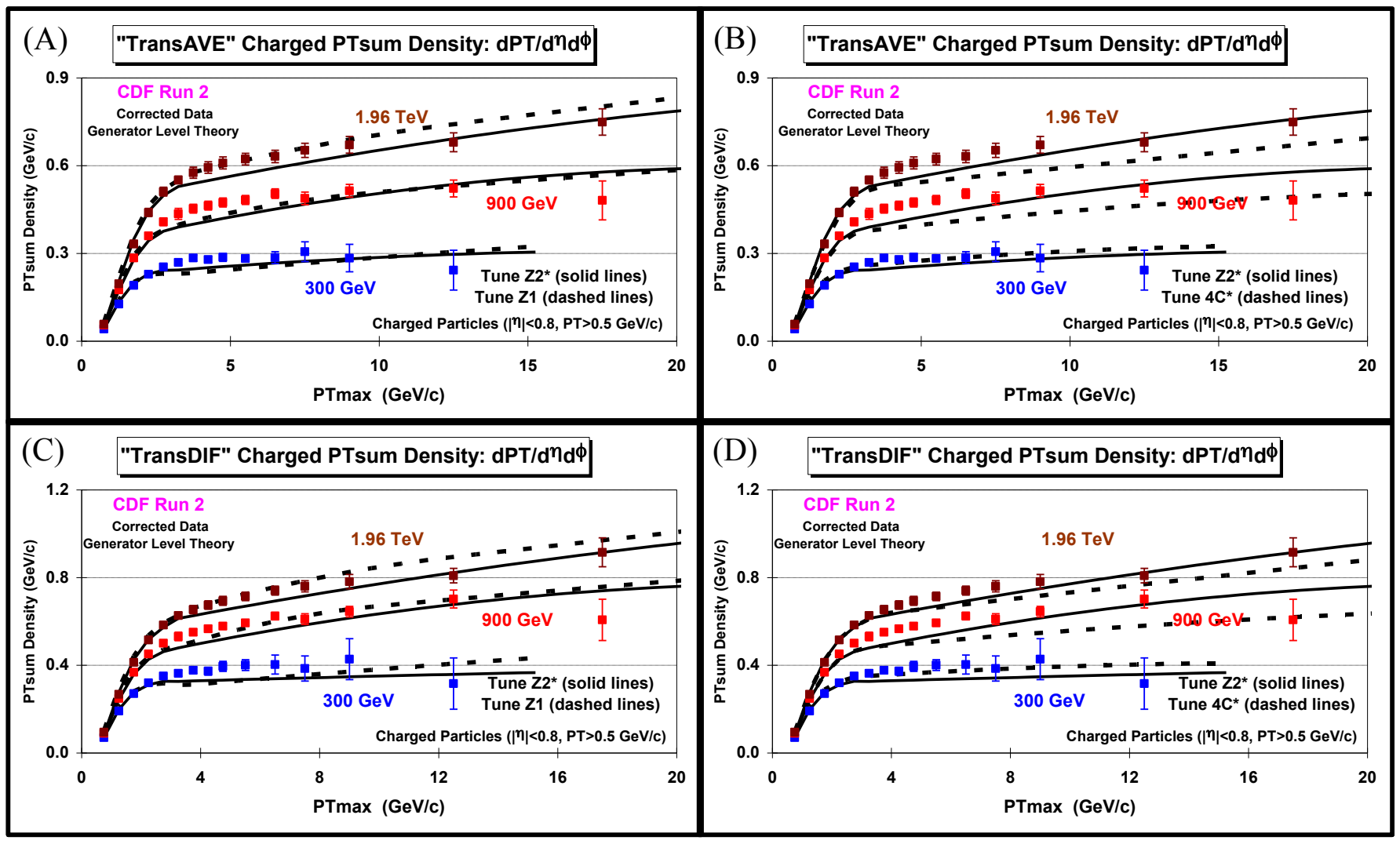

**Figure 13**. Data at 1.96 TeV, 900 GeV, and 300 GeV on the charged PTsum density ($p_T$ > 0.5 GeV/c, |η| < 0.8) in the transAVE **(A,B)** and transDIF **(C,D)** regions as defined by the leading charged particle, as a function of the $p_T$ of the leading charged particle, PTmax. The data are corrected to the particle level with errors that include both the statistical error and the systematic uncertainty, and are compared with PYTHIA Tune Z1 and Z2* **(A,C)** and PYTHIA Tune Z2* and 4C* **(B,D)**.

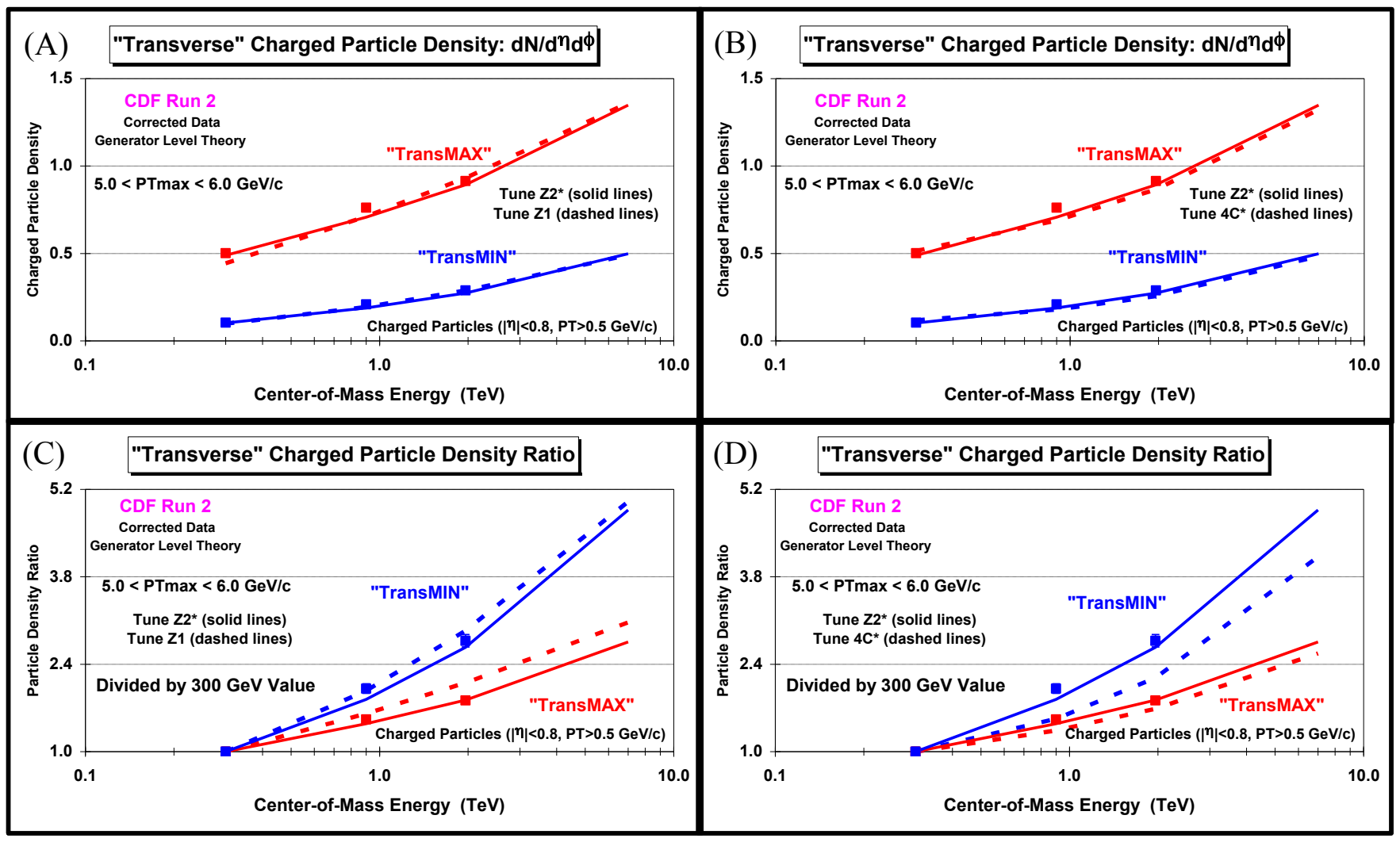

**Figure 14**. **(A,B)** Data on the transMAX and transMIN charged particle density as defined by the leading charged particle, for 5.0 < PTmax < 6.0 GeV/c plotted versus the center-of-mass energy for charged particles with $p_T$ > 0.5 GeV/c and |η| < 0.8. **(C,D)** Ratio of the data at 1.96 TeV, 900 GeV, and 300 GeV to the corresponding value at 300 GeV for the transMAX and transMIN charged particle density plotted versus the center-of-mass energy. The data are corrected to the particle level with errors that include both the statistical error and the systematic uncertainty, and are compared with PYTHIA Tune Z1 and Z2* **(A,C)** and PYTHIA Tune Z2* and 4C* **(B,D)**. The theory curves have been extrapolated to 7 TeV.

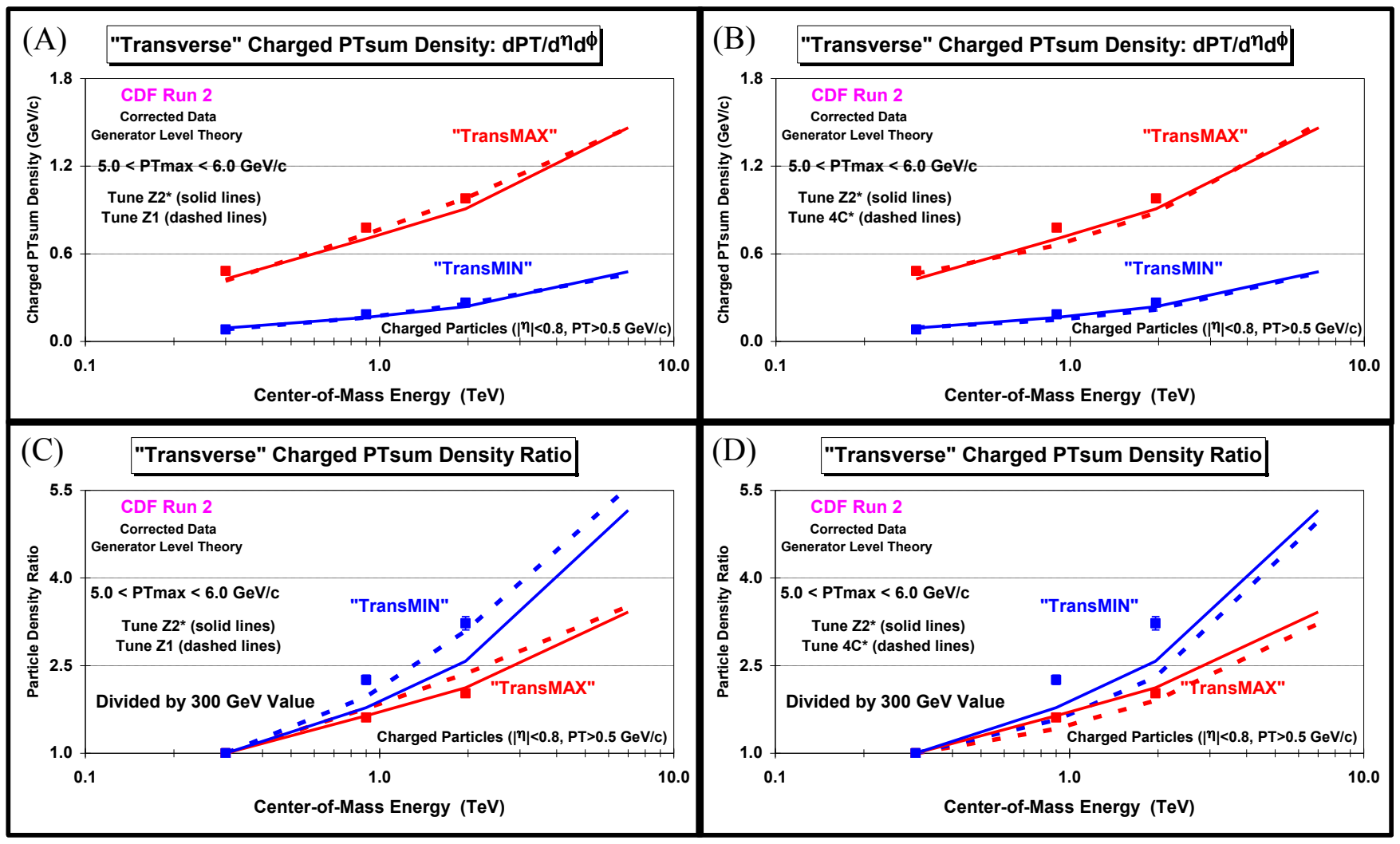

**Figure 15**. **(A,B)** Data on the transMAX and transMIN charged PTsum density as defined by the leading charged particle, for 5.0 < PTmax < 6.0 GeV/c plotted versus the center-of-mass energy for charged particles with $p_T$ > 0.5 GeV/c and |η| < 0.8. **(C,D)** Ratio of the data at 1.96 TeV, 900 GeV, and 300 GeV to the corresponding value at 300 GeV plotted versus the center-of-mass energy. The data are corrected to the particle level with errors that include both the statistical error and the systematic uncertainty, and are compared with PYTHIA Tune Z1 and Z2* **(A,C)** and PYTHIA Tune Z2* and 4C* **(B,D)**. The theory curves have been extrapolated to 7 TeV.

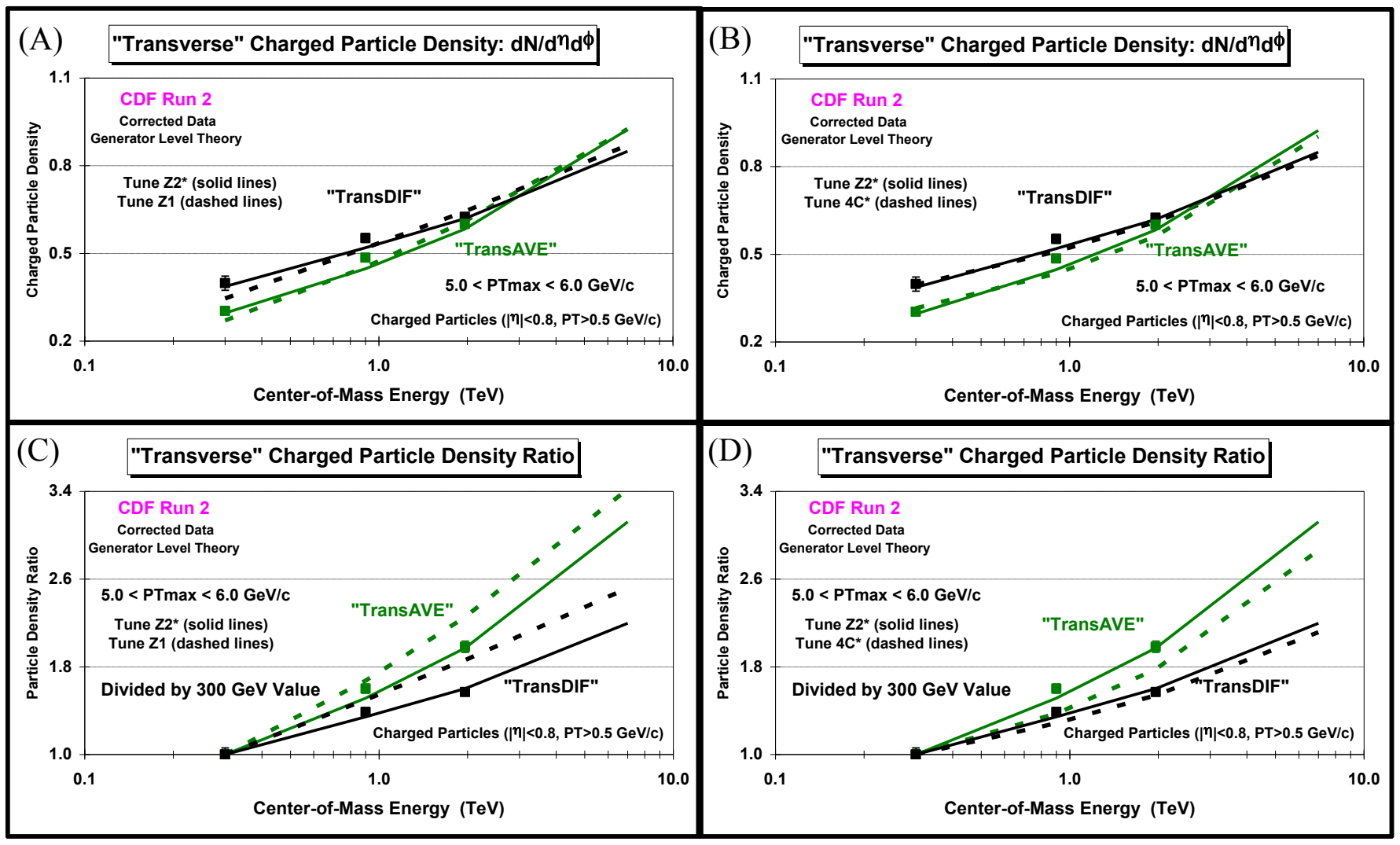

**Figure 16**. **(A,B)** Data on the transAVE and transDIF charged particle density as defined by the leading charged particle, for 5.0 < PTmax < 6.0 GeV/c plotted versus the center-of-mass energy for charged particles with $p_T$ > 0.5 GeV/c and $|\eta| < 0.8$. **(C,D)** Ratio of the data at 1.96 TeV, 900 GeV, and 300 GeV to the corresponding value at 300 GeV plotted versus the center-of-mass energy. The data are corrected to the particle level with errors that include both the statistical error and the systematic uncertainty, and are compared with PYTHIA Tune Z1 and Z2* **(A,C)** and PYTHIA Tune Z2* and 4C* **(B,D)**. The theory curves have been extrapolated to 7 TeV.

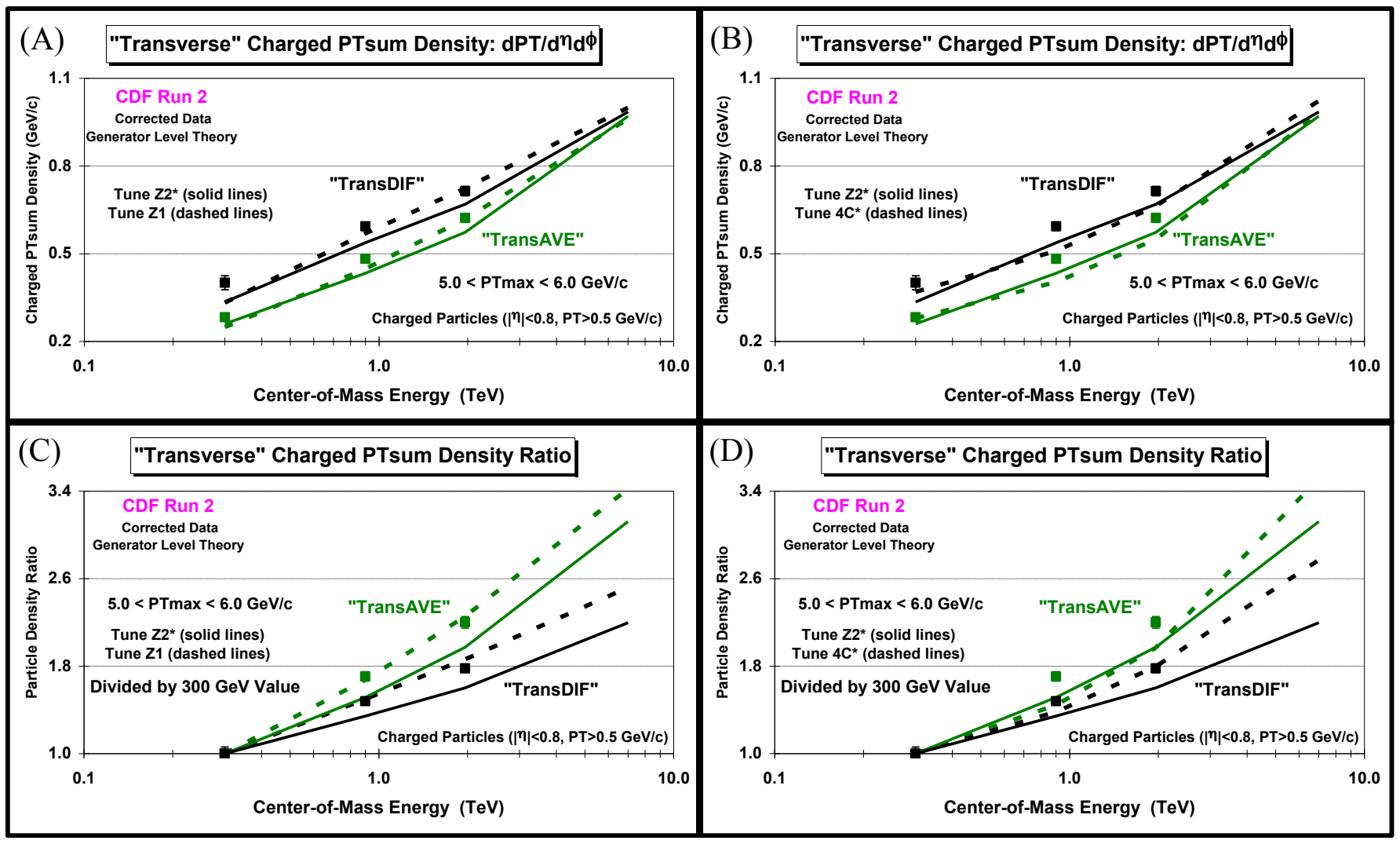

**Figure 17**. **(A,B)** Data on the transAVE and transDIF charged PTsum density as defined by the leading charged particle, for 5.0 < PTmax < 6.0 GeV/c plotted versus the center-of-mass energy for charged particles with $p_T$ > 0.5 GeV/c and $|\eta| < 0.8$. **(C,D)** Ratio of the data at 1.96 TeV, 900 GeV, and 300 GeV to the corresponding value at 300 GeV plotted versus the center-of-mass energy. The data are corrected to the particle level with errors that include both the statistical error and the systematic uncertainty, and are compared with PYTHIA Tune Z1 and Z2* **(A,C)** and PYTHIA Tune Z2* and 4C* **(B,D)**. The theory curves have been extrapolated to 7 TeV.

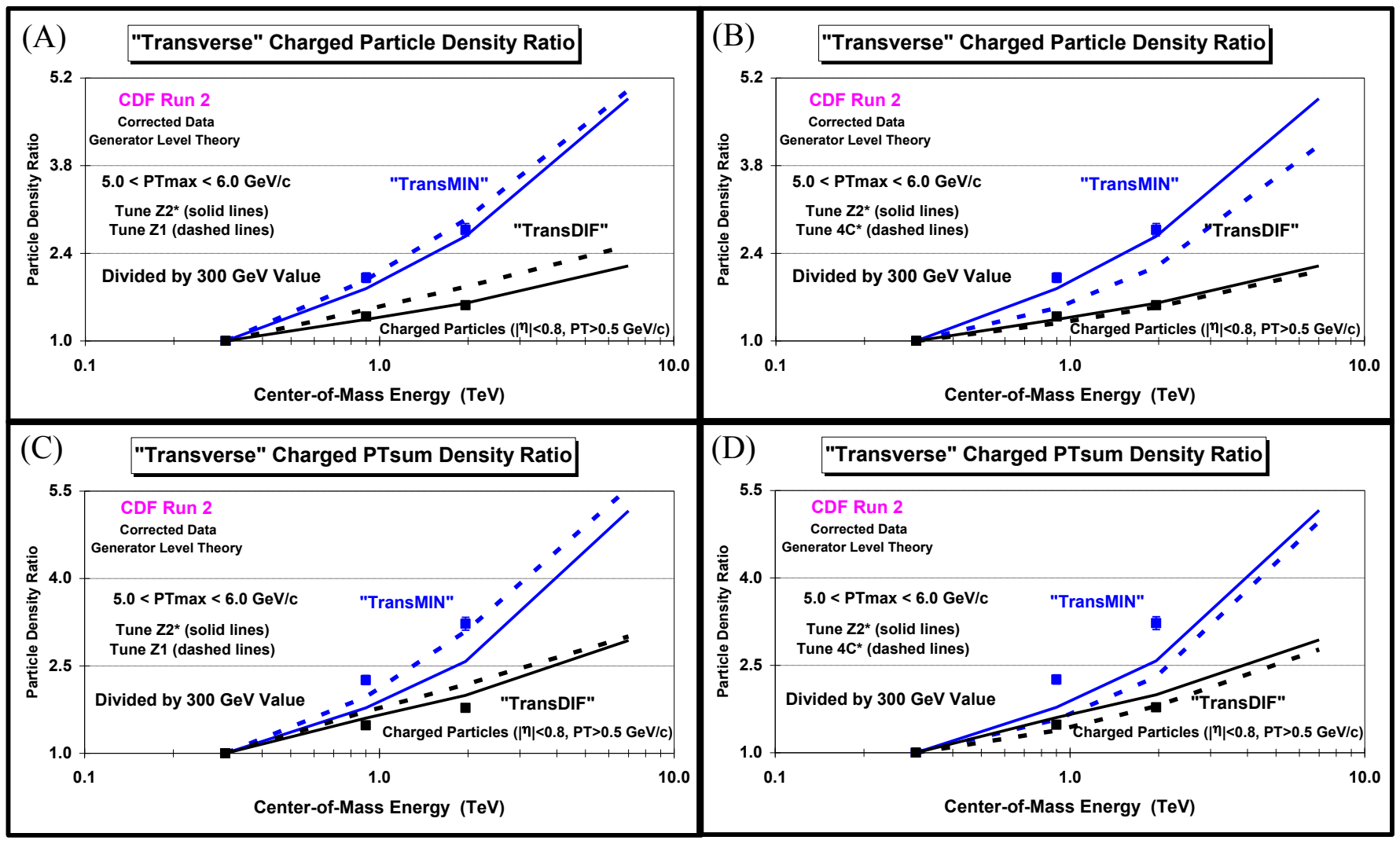

**Figure 18**. Ratio of the data at 1.96 TeV, 900 GeV, and 300 GeV to the corresponding value at 300 GeV for the transMIN and transDIF charged particle density **(A,B)** and charged PTsum density **(C,D)** as defined by the leading charged particle, for 5.0 < PTmax < 6.0 GeV/c plotted versus the center-of-mass energy. The data are corrected to the particle level with errors that include both the statistical error and the systematic uncertainty, and are compared with PYTHIA Tune Z1 and Z2* **(A,C)** and PYTHIA Tune Z2* and 4C* **(B,D)**. The theory curves have been extrapolated to 7 TeV.

Figure 18 compares the energy dependence of the transMIN and transDIF components. The data show that the transMIN charged particle and charged PTsum density increase by a factor of 2.8 and 3.2, respectively, in going from 300 GeV to 1.96 TeV, while the transDIF charged particle and charged PTsum density increases by only a factor of 1.6 and 1.8, respectively. The transMIN density (more sensitive to MPI & BBR) increases much faster with center-of-mass energy than does the transDIF density (more sensitive to ISR & FSR). The MPI increases like a power of the center-of-mass energy (or a power of the log of the energy), while the ISR & FSR increase logarithmically. This is the first time we have seen the different energy dependences of these two components. Previously we only had information on the energy dependence of the transAVE density. The QCD Monte Carlo tunes do a fairly good (although not perfect) job in describing the energy dependence of transMIN and transDIF.

## (4) The Transverse Average $P_T$

Figure 19 shows the data at 1.96 TeV, 900 GeV, and 300 GeV on the charged-particle average $p_T$ in the transverse region as defined by the leading charged particle, as a function of the $p_T$ of the leading charged particle, PTmax. Figure 19 also shows the transverse charged particle average $p_T$ for 5.0 < PTmax < 6.0 GeV/c plotted versus the center-of-mass energy. The transverse average $p_T$ increases slowly with center-of-mass energy and this slow rise is correctly predicted by the QCD Monte Carlo model tunes. However, all the tunes predict an average $p_T$ that is slightly less than that seen in the data over most of the PTmax range. The average $p_T$ is a measure to the $p_T$ distribution of charged particles and the tunes predict a $p_T$ distribution that is slightly too soft.

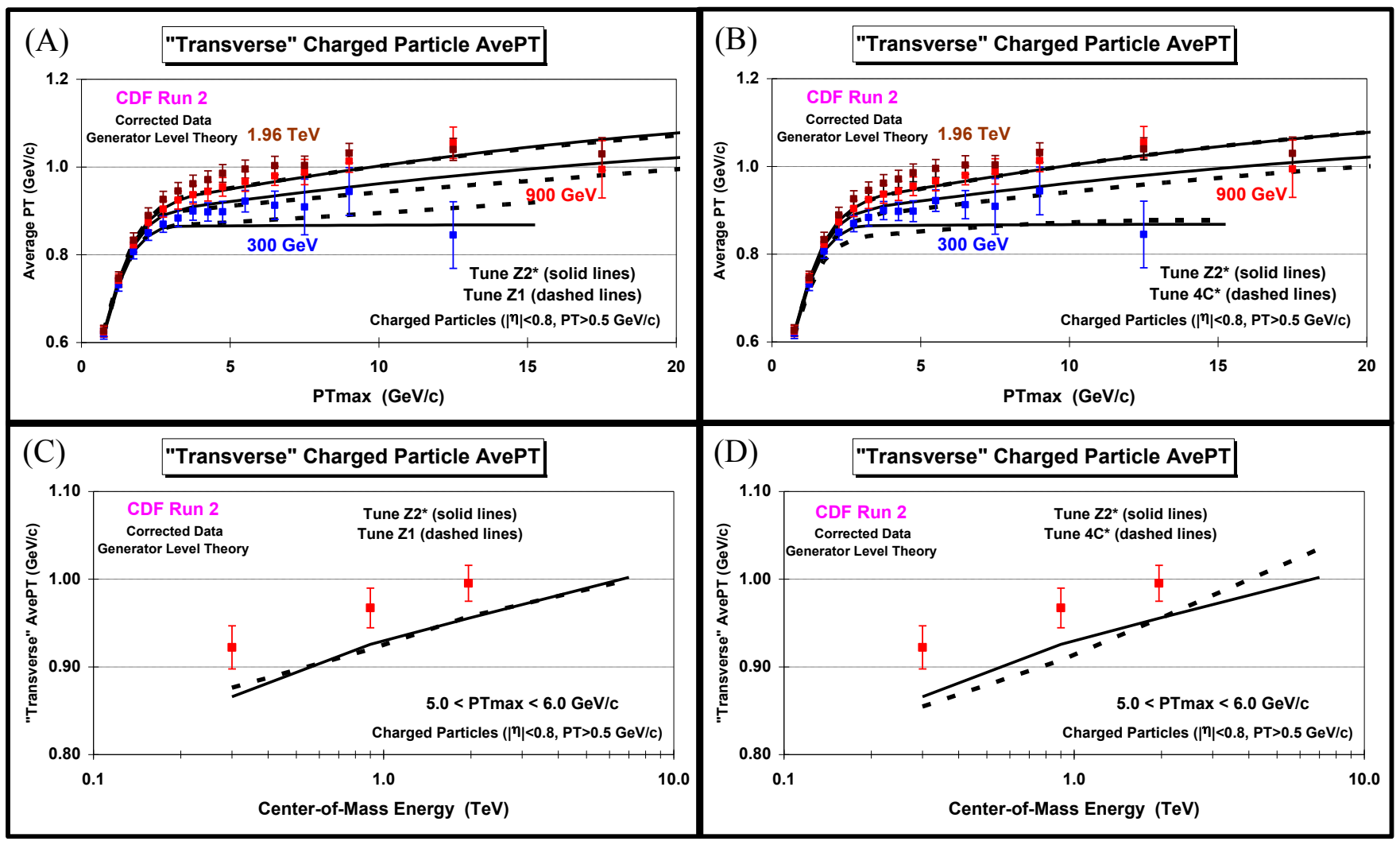

**Figure 19**. **(A,B)** Data at 1.96 TeV, 900 GeV, and 300 GeV on the charged particle average $p_T$ ($p_T > 0.5$ GeV/c, $|\eta| < 0.8$) in the transverse region as defined by the leading charged particle, as a function of the $p_T$ of the leading charged particle, PTmax. **(C,D)** Data on the transverse charged particle average $p_T$ as defined by the leading charged particle, for 5.0 < PTmax < 6.0 GeV/c plotted versus the center-of-mass energy for charged particles with $p_T > 0.5$ GeV/c and $|\eta| < 0.8$. The data are corrected to the particle level with errors that include both the statistical error and the systematic uncertainty, and are compared with PYTHIA Tune Z1 and Z2* **(A,C)** and PYTHIA Tune Z2* and 4C* **(B,D)**.

# IV. Summary and Conclusions

We first examine the average overall total number of charged particles and the pseudorapidity distribution of charged particles at 300 GeV, 900 GeV, and 1.96 TeV. We then show how the average overall number of charged particles depends on the center-of-mass energy, and the transverse momentum of the leading charged particle, PTmax. The QCD Monte Carlo model tunes do a fairly good job predicting the correct overall number of charged particles at the three energies and they correctly describe how the overall number of charged particles depends on PTmax. In addition, we study the associated charged particle and charged PTsum density. The leading charged particle is not included in the associated density. The QCD Monte Carlo model tunes describe the overall associated densities fairly well, however, at 1.96 TeV and 900 GeV the tunes produce slightly too many associated charged particles at large PTmax values.

To study the event topology, the associated density is divided into the toward, away, and transverse (*i.e.,* transAVE) densities. As PTmax increases the toward-side and away-side charged particle and PTsum densities become much larger than they are in the transverse region, since they typically receive significant contributions from the two leading hard-scattered jets. At large PTmax values at 300 GeV the charged-particle and PTsum densities are larger in the away region than they are in the toward region. At 900 GeV they are roughly the same, and at 1.96 TeV the densities in the toward region are larger than they are in the away region. The PYTHIA tunes do a good job describing the topological structure of the event. There is a tendency for the tunes to produce too much associated density in the toward region, something we saw in the first CDF underlying event analysis in 2002 [2].

To study the underlying event (UE) in more detail, the two transverse regions are distinguished as a transMAX region and a transMIN region and we compare the center-of-mass energy dependence of the transMIN and transDIF densities. The transverse (*i.e.,* transAVE) density is the average of the transMAX and transMIN densities, while transDIF is the transMAX density minus the transMIN density. The transMIN densities are sensitive to the modeling of the multiple parton interactions (MPI) and beam-beam remnant (BBR) components of the UE, while the transDIF densities are sensitive to initial-state and final-state radiation (ISR & FSR). The data show that the transMIN charged-particle and charged-PTsum densities increase by a factor of 2.8 and 3.2, respectively, in going from 300 GeV to 1.96 TeV, while the transDIF charged-particle and charged-PTsum densities increases by only a factor of 1.6 and 1.8, respectively. The transMIN densities increase much faster with center-of-mass energy than do the transDIF densities. The MPI increases like a power of the center-of-mass energy (or the log of the energy to a power), while the ISR & FSR increase logarithmically. This is the first time we have seen the different energy dependences of these two components. Previously, we only had information on the energy dependence of the transAVE density. The QCD Monte Carlo model tunes describe fairly well the energy dependence of the transMIN and transDIF densities.

On must have UE data at a minimum of three center-of-mass energies to test the energy dependence of the QCD Monte Carlo models. The PYTHIA 6.4 Tune Z1 and Z2* and the PYTHIA 8 Tune 4C* do a nice job in describing the LHC UE data at 7 TeV [10]. They also describe fairly well all of the general features of the CDF data at 300 GeV, 900 GeV, and 1.96 TeV. The data presented here provide the first true test of the ability of the QCD Monte Carlo models to describe the energy dependence of the UE in hadron-hadron collisions. The PYTHIA tunes do a fairly good job in describing the data, although they do not describe the data perfectly. Combining the CDF data from the Tevatron Energy Scan presented here with LHC data at 7 TeV will allow for detailed studies of the energy dependence of hadron-hadron collisions, which will improve the QCD Monte Carlo model tunes, resulting in more precise predictions at the LHC energies of 13 and 14 TeV.

## ACKNOWLEDGMENTS

We thank the Fermilab staff and the technical staffs of the participating institutions for their vital contributions. This work was supported by the U.S. Department of Energy and National Science Foundation; the Italian Istituto Nazionale di Fisica Nucleare; the Ministry of Education, Culture, Sports, Science and Technology of Japan; the Natural Sciences and Engineering Research Council of Canada; the National Science Council of the Republic of China; the Swiss National Science Foundation; the A.P. Sloan Foundation; the Bundesministerium für Bildung und Forschung, Germany; the Korean World Class University Program, the National Research Foundation of Korea; the Science and Technology Facilities Council and the Royal Society, United Kingdom; the Russian Foundation for Basic Research; the Ministerio de Ciencia e Innovación, and Programa Consolider-Ingenio 2010, Spain; the Slovak R\&D Agency; the Academy of Finland; the Australian Research Council (ARC); and the EU community Marie Curie Fellowship Contract No. 302103.

## References

[1] R. Field, Annual Review of Nuclear and Particle Science, **62**, 427–457 (2012).

[2] T. Aaltonen et al. (CDF Collaboration), Phys. Rev. **D65**, 092002, (2002).

[3] T. Aaltonen et al. (CDF Collaboration), Phys. Rev. **D82**, 034001 (2010), arXiv:1003.3146.

[4] Using transMAX and transMIN was first suggested by Bryan Webber and implemented in a paper by Jon Pumplin, Phys. Rev. **D57**, 5787 (1998).

[5] T. Sjöstrand, Phys. Lett. **157B**, 321 (1985); M. Bengtsson, T. Sjöstrand, and M. van Zijl, Z. Phys. **C32**, 67 (1986); T. Sjöstrand and M. van Zijl, Phys. Rev. **D36**, 2019 (1987). T. Sjöstrand, P. Eden, C. Friberg, L. Lonnblad, G. Miu, S. Mrenna and E. Norrbin, Computer Physics Commun. **135**, 238 (2001).

[6] R. Field, Tevatron-for-LHC: Report of the QCD Working Group, arXiv:hep-ph/0610012, FERMILAB-Conf-06-359, October 1, 2006.

[7] R. Field, proceedings of the First International Workshop on Multiple Partonic Interactions at the LHC (MPI08), Perugia, Italy, October, 2009. arXiv:1003.4220.

[8] P. Skands, arXiv:0905.3418.

[9] R. Field, arXiv:1010.3558, proceedings of the Hadron Collider Physics Symposium (HCP2010), August 23-27, 2010.

[10] R. Field, arXiv:1110.5530, proceedings of the 51st Cracow School of Theoretical Physics: *The Soft Side of the LHC*, Zakopane, June 11 - 19, 2011, Acta Physica Polonica **B42**, 2631 (2011).

[11] Serguei Chatrchyan et al. (CMS Collaboration), J. High Energy Phys. **09** (2011) 109, arXiv:1107.0330.

[12] H. L. Lai *et al.* (CTEQ Collaboration), Eur. Phys. J. **C12**, 375-392 (2000).

[13] T. Sjöstrand, S. Mrenna, and P. Skands, A Brief Introduction to PYTHIA 8.1, Comput. Phys. Commun. **178**, 852–867 (2008), arXiv:0710.3820 [hep-ph].

[14] R. Corke, T. Sjöstrand, J. High Energy Phys. **1103**:032 (2011), arXiv:1011.1759 [hep-ph].

[15] D. Acosta *et al.* (CDF Collaboration), Phys. Rev. D **71**, 032001 (2005); D. Acosta *et al.* (CDF Collaboration), Phys. Rev. D **71**, 052003 (2005); A. Abulencia *et al.* (CDF Collaboration), J. Phys. G Nucl. Part. Phys. **34**, 2457 (2007).

[16] A. Affolder *et al.* (CDF Collaboration), Nucl. Instrum. Methods **A526**, 249 (2004).

[17] R. Brun *et al.* (1987), unpublished, CERN-DD/EE/84-1.

[18] G. Grindhammer, M. Rudowicz, and S. Peters, Nucl. Instrum. Methods **290**, 469 (1990).

[19] These plots were suggested by the MB&UE working group at the LHC Physics Center at CERN. See the LPCC website at http://lpcc.web.cern.ch/LPCC/.